# Electrodeposition of Ni–P composite coatings: A review

Aleksandra Lelevic[a],[*], Frank C. Walsh[b]

[a] *Artia Nanoengineering & Consulting, Athens, Greece*
[b] *Electrochemical Engineering Laboratory, National Centre for Advanced Tribology, Department of Mechanical Engineering, University of Southampton, Southampton SO17 1BJ, UK*

ABSTRACT

Ni–P coatings produced by electrodeposition have important mechanical, tribological and electrochemical properties. They can also exhibit catalytic activity and beneficial magnetic behaviour. With subsequent thermal treatment, the hardness of such Ni–P coatings can approach or exceed that of hard Cr coatings. Electrochemical codeposition of homogeneously dispersed second phase particles within the Ni–P matrix can enhance deposit properties and meet the challenging demands on modern engineering coatings. A general overview of research work on the electrodeposition of Ni–P composite coatings containing included ceramic or solid lubricant particles is provided. Advances in research into Ni–P composite layers reinforced by SiC, $B_4C$, WC, $Al_2O_3$, $SiO_2$, $TiO_2$, $CeO_2$, MWCNT, $MoS_2$, $WS_2$, TiN, hexagonal BN, PTFE and their combinations are considered. Major models proposed for the codeposition of particles, the influence of bath hydrodynamics and control of operational parameters are illustrated by examples. Important trends are highlighted and opportunities for future R & D are summarised.

## 1. Introduction

Ni–P has attracted considerable attention as an engineering coating [1]. Alloying nickel with phosphorus brings improvement in mechanical properties, wear and corrosion resistance, magnetic behaviour, a higher fatigue limit and lower macroscopic deformation [2]. Many studies have been conducted to show that coatings based on Ni–P, with careful tailoring of the composition and structure, can offer smart and adaptive solutions to a wide range of environmental conditions [3,4]. By applying subsequent thermal treatment the hardness of Ni–P electrodeposits can approach or surpass that of hard Cr coatings [5]. However, brittleness and reduced integrity of Ni–P electrodeposits, due to formation of intermetallic compounds during heat treatment, restricts their application to Cr replacement coatings. This is especially true for anti-wear applications involving high thickness deposits in a harsh environment, such as under the conditions of high speed and heavy load [6]. Electroplated Ni–P is also characterized by higher internal stress when compared to pure Ni electrodeposits [2].

The timeline in Fig. 1 summarises developments in electrodeposition of metal matrix composite coatings having particles as second phase materials, Ni-P-X (X = particle) composite systems and models describing the effect of operational conditions on the deposit composition. Following optimisation, the electrolytic codeposition of inert particles within the Ni–P matrix can contribute to the improvement of mechanical, tribological and electrochemical features inherent to the matrix [7–10]. Composite materials fabricated in such a way possess basic properties that originate from the matrix while the incorporated particles can enhance and/or add particular features or functionalities. The benefits attributable to the presence of particles dispersed within the metal matrix, depend on their content, degree of dispersion, material and shape. Two types of included particles have been employed to produce electroplated Ni–P composites, namely solid lubricant particles and hard, ceramic particles providing hardness, wear resistance and improved ability to act as a cutting tool. Regarding the first type of reinforcement, this paper covers the research work concerning the codeposition of $MoS_2$, $WS_2$, CNT, carbon black, PTFE or hexagonal BN; from the latter group, Ni–P composites containing SiC, $B_4C$, WC, $Al_2O_3$, $CeO_2$, $SiO_2$, $TiO_2$ or TiN are considered. A review by Berkh and Zahavi [4] provides more information on electrodeposition of Ni–P composites containing B, TiC, $CaF_2$, $Si_3N_4$, $ZrO_2$, $Fe_{80}B_{20}$ and $HfB_2$, with particle

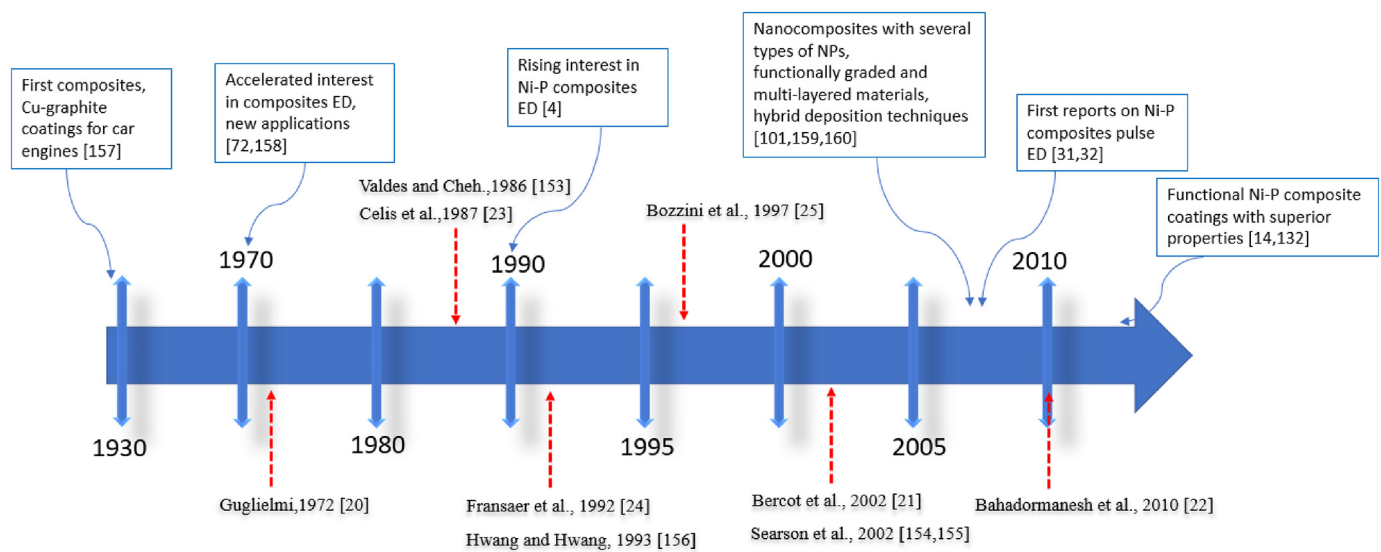

**Fig. 1.** A timeline illustrating developments in Ni—particle electroplating, considering a) composite electrodeposition[72,101,157–160], b) Ni-P electrodeposition [4,14,31,32,132] and c) models describing the effect of operating conditions on the deposit[20–25,153–156]. Red arrows indicate research directions in terms of proposed models for codeposition of particles, blue arrows illustrate trends in metallic matrix composites electroplating, with a focus on Ni and Ni-P based deposits. (For interpretation of the references to colour in this figure legend, the reader is referred to the web version of this article.)

sizes ranging from 0.2 to 7 μm.

## 2. Composite electrodeposition

There are many benefits of incorporating inert particles into a metallic matrix via electroplating and many operational factors must be considered (Fig. 2). Normally, an undivided cell is used with a soluble nickel anode in an uncomplexed acidic electrolyte containing reinforcement particles.

Using a soluble anode, nickel ions, formed via (Eq. (1))

$$Ni = Ni^{2+} + 2e^-  \quad (1)$$
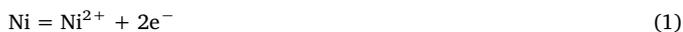

replenish those lost by deposition of nickel at the cathode.

At an insoluble anode, such as platinum in the laboratory or platinised titanium, in industry, oxygen evolution takes place (Eq. (2)).

$$2H_2O = O_2 + 4H^+ + 4e^- \quad (2)$$
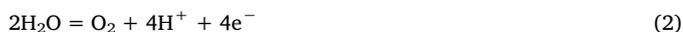

At the cathode workpiece, the primary reaction of nickel deposition (Eq. (3))

$$Ni^{2+} + 2e^- = Ni \quad (3)$$
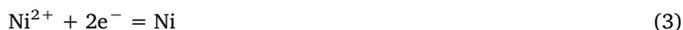

is accompanied by deposition of phosphorus. Phosphorus can be produced by reduction of hypophosphite ions, e.g. (Eq. (4))

$$H_2PO_3^- + 4H^+ + 3e^- = P + 3H_2O \quad (4)$$
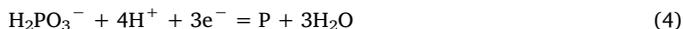

Indirect reactions are also possible [1].

Hydrogen evolution also occurs at the cathode surface (Eq. (5))

$$2H^+ + 2e^- = H_2 \quad (5)$$
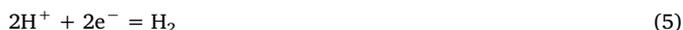

This not only lowers current efficiency and increases local pH but can result in deposit porosity due to hydrogen gas bubbles sticking to the surface, if the electrolyte/electrode movement is limited or insufficient surfactant is present in the bath to promote surface wetting and bubble release.

Inclusion and uniform distribution of nanometer or micrometer sized second phase particles can lead to improvement of inherent mechanical, tribological, electrochemical properties of the material and open pathways into completely new material applications. Particles can be suspended in the electrolyte by cathode movement, electrolyte agitation or the use of surfactants. Particles in the bath develop a natural surface charge and zeta potential. This surface charge can be modified in sign and size through adsorption of ions from metal source salts but also by the use of surfactant additions to the bath. For example, the use of a strongly cationic surfactant can result in a significant positive charge on the surface of the particles, aiding migration to the cathode. In a simple case, a cationic surfactant, such as CTAB can adsorb on the surface of the particle, conferring a positive charge. This charge facilitated electrophoretic transport to the cathode, along with convective-diffusion of particles, which depends on relative cathode/electrolyte movement and mass transport [10]. Despite their importance, the quality and stability of particle dispersion in the bath and choice of surfactant have been insufficiently studied. The surfactant concentration in the bath is critical, as an excess can result in deleterious effects on the deposit, including high internal stress and brittleness. It is surprising how few studies have screened surfactants [11] and characterized the stability of particle dispersion in the bath prior to electrodeposition [12].

Table 1 gives an overview of the type of reinforcement used to deposit several Ni—P composites via electroplating, with the particle sizes, proposed bath loads, resultant deposits and their major features, while Fig. 3 indicates the major types of inclusion and their application areas.

A number of studies have demonstrated that the use of nanosized particles rather than the sub-micron or micron-size ones can give rise to more substantial improvement of the deposit's properties [10]. Uniformly dispersed nanoparticles in the deposit contribute to significant strengthening of the material (dispersion strengthening) which can be preserved even at higher temperatures owing to thermal stability of the reinforcing particles. According to Zhang and Chen [13] the strengthening effects in particulate-reinforced metal matrix nanocomposites arise from Orowan strengthening, enhanced dislocation density generated by the difference in the coefficient of thermal expansion between the matrix and the particles and load-bearing effect. Particles contribute to significant grain refinement as their presence provides more sites for nucleation and retards the growth of crystals producing a smaller grain size [14]. With the increasing availability of nanoparticles, interest in electroplating nanocomposites continues to grow. Major challenges are achieving high codeposition rates of particles, their homogeneous distribution in the metal matrix and producing consistent deposits having controlled composition and properties.

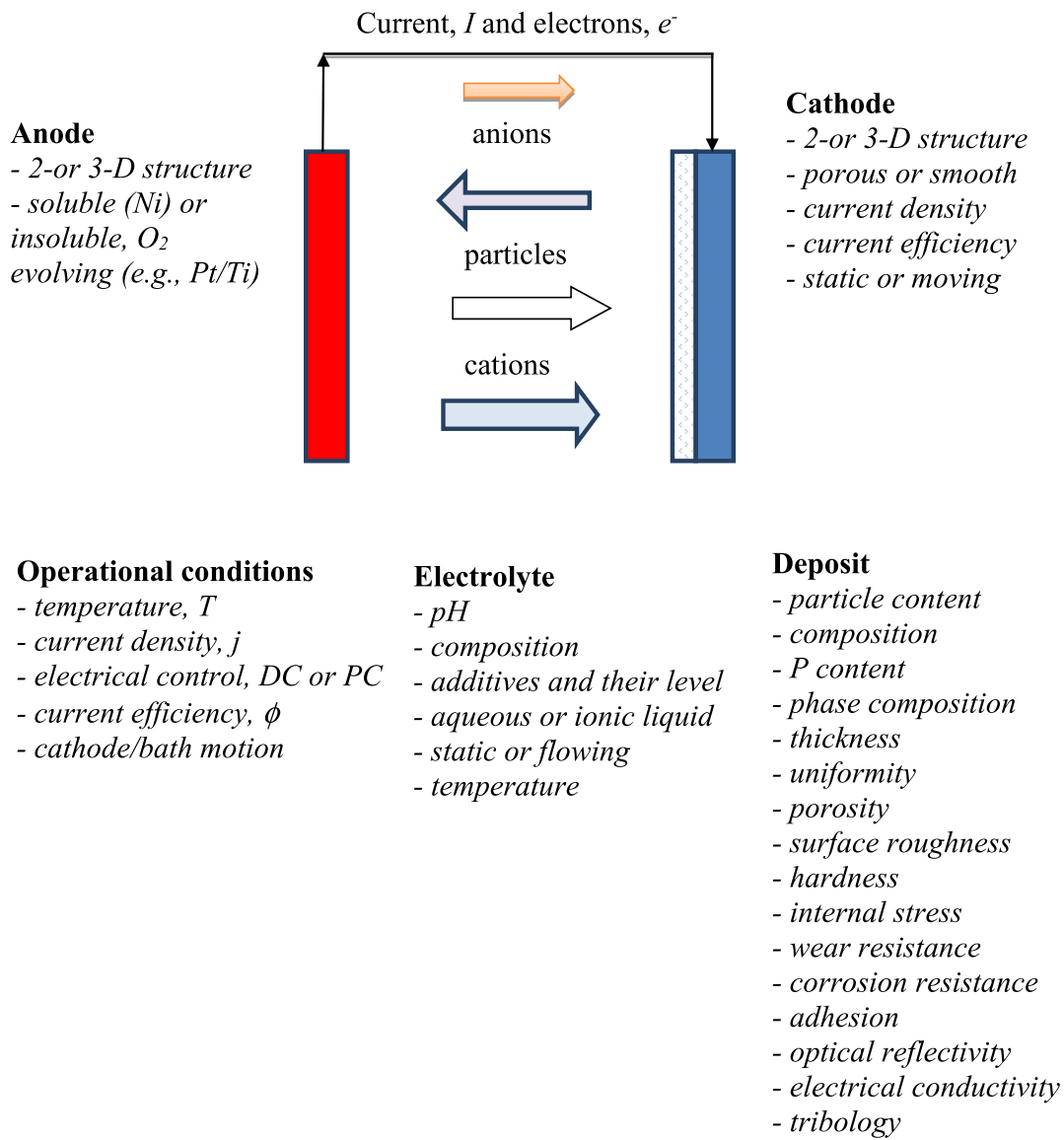

Fig. 2. A schematic diagram of an undivided cell for electrodeposition of a nickel-phosphorus-particle composite coating. Important aspects of electrolyte choice and control of operational conditions are indicated.

Table 1
Different types of included particles and their recommended bath loads for optimal enhancement of Ni-P composite functional properties (DC—direct current, PC—pulse current, WCA—water contact angle, COF—coefficient of friction, $d$—diameter, $l$—length).

| Reference | Reinforcement type | Particle size | Bath load | Substrate | Content in deposit (wt%) | P content | As-plated deposit properties |
|---|---|---|---|---|---|---|---|
| [32] | SiC | 300 nm | 20 g L$^{-1}$ | Mild steel | 1.5 (PC) | ~4 wt% | Hard coating (~700 HV); > 50 μm |
| [31] | SiC | 1 μm | 20 g L$^{-1}$ | Brass | 22 (PC) | Low phosphorus | Hard coating (~730 HV); > 50 μm |
| [84] | WC | 200 nm | 20 g L$^{-1}$ | Brass | > 30 (PC) | ~13 wt% | Hard coating (~700 HV); ~40 μm |
| [83] | B$_4$C | 1.2 μm | 5 g L$^{-1}$ | Gold; iron | / (PC) | 4 and 12 wt% | Hard coating (~830 HV); ~50 μm |
| [91] | TiO$_2$ | 5–15 μm | 200 g L$^{-1}$ | Copper | 24.2 (DC) | 23.2 wt% | Electroactive material; ~25 μm |
| [106] | CeO$_2$ | 20–50 nm | 15 g L$^{-1}$ | Iron | 2.3 (PC) | 8.6 wt% | Hard coating (~575 HV); 50 μm |
| [132] | MoS$_2$ | 1–4 μm | 10 g L$^{-1}$ | Mild steel | 7.9 (DC) | Low phosphorus | Self-lubricating (COF 0.05) |
| [14] | WS$_2$ | 100–300 nm | 15 g L$^{-1}$ | Mild steel | 4.8 (DC) | Low phosphorus | Hydrophobic (WCA 157°), COF 0.17; < 40 μm |
| [146] | MWCNT | $d$ = 100–200 nm $l$ = 20 μm | 2 g L$^{-1}$ | Copper; stainless steel | 0.7 (DC) | 20–22 at.% | Self-lubricating (COF 0.1–0.2); ~200 μm |
| [86] | TiN | 20 nm | 5 g L$^{-1}$ | Mg alloy | (~0.5 (DC)) | High phosphorus | Corrosion resistant (−0.4 V vs. SCE > 1600 h); ~10 μm |

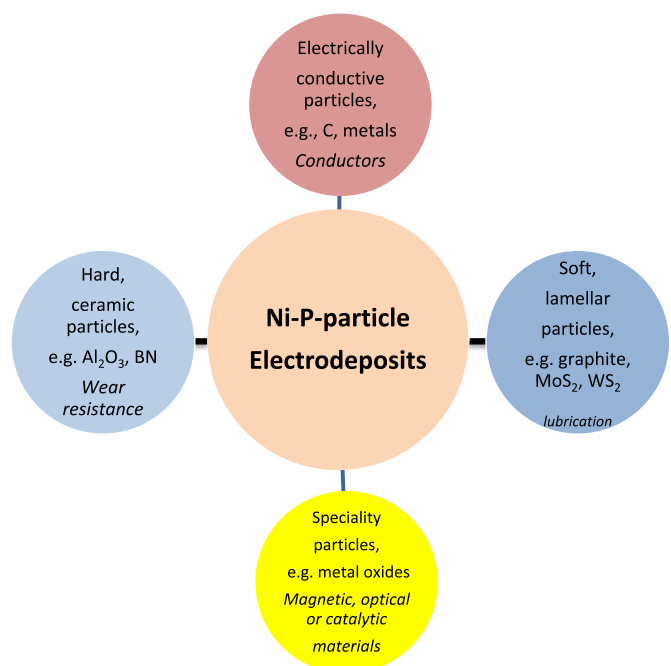

**Fig. 3.** A simple classification of the major types of particle inclusions electrodeposited in Ni-P-particle composite coatings and their major fields of application.

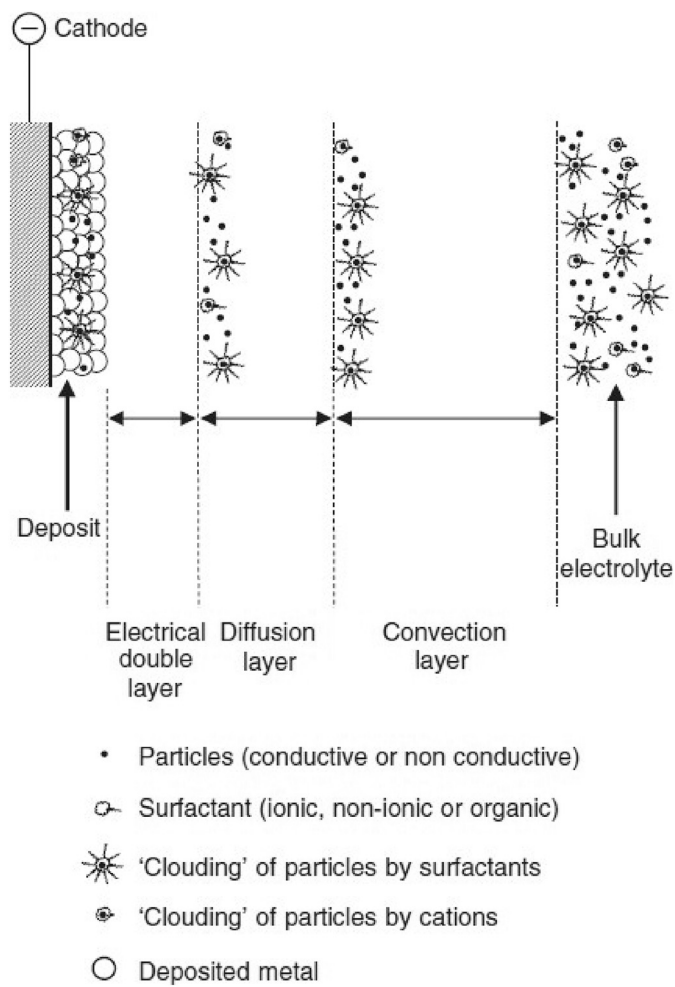

**Fig. 4.** Particle codeposition in a metal deposit.
Reprinted with permission from [16]. Copyright (2006) Elsevier.

Improvements in deposit characteristics are strongly dependent on the type of included particle. Hard, ceramic particles can contribute to mechanical strength and wear resistance while soft, lubricant particles can improve tribological performance. Metallic matrix based coatings comprising both hard and lubricant particles can also be selected. In such cases, lubricant/soft phase provides lubrication to a surface, hard ceramic phase provides structural integrity and wear resistance of the coating and a ductile metal phase ensures ductility and matrix support. A non-line-of-sight ("NLOS") process for electroplating lubricant-hard-ductile nanocomposite coatings has been demonstrated in [15].

The codeposition of particles in a metallic matrix during electroplating occurs in several stages, including formation of ionic clouds around the particles in the bulk electrolyte, convection towards the cathode, diffusion through a hydrodynamic boundary layer, diffusion through a concentration boundary layer and finally adsorption at the cathode surface where particles are entrapped within the metal deposit (Fig. 4) [16]. Incorporation of particles within the deposit depends on many process parameters including the characteristics of the particles (e.g. loading, surface charge, type, shape and size), electrolyte composition (e.g. electrolyte concentration, additives, temperature, pH, surfactant type and its concentration), applied current density (e.g. direct current, pulsed current, pulse time, duty cycle, potentiostatic versus galvanostatic control), flow environment inside the electroplating tank (e.g. laminar, mixed, turbulent regime) and shape/size of electroplating tank and electrode geometry (rotating disk electrode, rotating cylinder electrode, plate-in-tanks, parallel plate electrodes and many variations of electroplating tanks) [16,17].

A number of models have been proposed to describe and rationalise the inclusion of particles within a metal matrix during electrodeposition and several review papers consider the main approaches used in these models [7,10,16,18,19].

Guglielmi [20] devised the first comprehensive model for codeposition of particles. The model assumed two adsorption steps, one being physical and the other electrochemical in nature. This enabled the effect of current density and particle loading on particle codeposition rate to be rationalised. Guglielmi's model presented a kernel for many subsequently proposed models and its validity has been verified in many studies employing different deposit compositions. The main drawback of Guglielmi's approach is the neglect of particle size effects, hydrodynamics and convective-diffusion mass transfer on the particle incorporation process. Modifications of this model have been proposed by several other research groups. Bercot et al. [21] additionally considered hydrodynamic conditions and the effects of surface adsorption and electrolyte flow. They devised a corrective factor for Guglielmi's model allowing to study the incorporation of PTFE particles into nickel electrolytic coatings performed under the conditions of magnetic stirring. Bahadormanesh and colleagues [22] modified the Guglielmi model in order to study the deposition of high-volume percentages of the second phase particles. The revised model was employed to describe the effects of current density, particle load and stirring rate of the electrolytic bath on the kinetics of codeposition.

Alternative models have been proposed. Celis et al. [23] considered that the particle will only be embedded in the growing deposit if a certain amount of adsorbed ions on its surface is reduced. They derived an equation allowing to predict the degree of codeposition at a given current density for a system of defined hydrodynamics (e.g. rotating disc electrode). In the Trajectory model devised by Fransaer and colleagues [24], the codeposition rate is determined by considering fluid flow in the electrolyte, taking into the account all forces acting on the particle and assuming that particles will be incorporated in the matrix once the contact with the electrode surface occurs. Bozzini et al. [25] studied the effect of suspended particle concentration in the bath on the volume fraction and the size distribution of the embedded particles in the deposit under fixed hydrodynamic conditions.

The complexity of composite electroplating makes it difficult to quantify the importance of bath and operating conditions, in order to achieve robust, generalised models applicable to large scale plating. General drawbacks of the presently advocated models are that they are specific to certain electrodeposition conditions, often ignore the impact of significant process influencers (different particle size and shapes, hydrodynamics, current density, etc.) with the occurring discrepancies in definition of certain process parameters [10].

The majority of recent investigations indicate that three global factors can be identified as influencing the particle codeposition, namely, applied current density, bath agitation or electrode movement and particle type, concentration and size [16]. The choice of particles and their properties is critical. Ehrhardt [26] found that electrically conductive SiC codeposited more easily in the Ni matrix, compared to non-conductive $Al_2O_3$. In the case of hydrophilic particles, a larger weight fraction in the deposit is obtained for conductive particles compared to the case of codeposition of non-conductive species, while in the case of low wettability particles lower particle content is obtained than in the event of employing wettable ones [4]. These results contradict those reported by Terzieva et al. who studied alumina particle incorporation into copper deposits [164]. Regarding particle size, Surviliene et al. [27] studied the codeposition of $MoO_2$ and $TiO_2$ particles with chromium. They found that the particle type had more influence on coating morphology when compared to the particle size. This finding is corroborated by the study of Chen et al. [28] who also observed that particle type exerts greater impact on the codeposition process. They found α-$Al_2O_3$ particles to be more easily codeposited with copper when compared to γ-$Al_2O_3$. A recent doctoral thesis [2] has summarised the effects of particle types and sizes in Ni and Ni−P composite plating. The importance of particle type was explored by employing $Al_2O_3$, $TiO_2$ and SiC nanoparticles having the same nominal size (~50 nm). The degree of particle incorporation, coating morphology, microstructure, and residual stress were very different for these incorporated particles. For Ni−P, codeposition of $TiO_2$ resulted in the highest incorporation of inclusions and the highest decline in residual stress after annealing. Regarding particle size, Sadeghi [2] investigated the codeposition of $Al_2O_3$ particles of nano and sub-micron size. Ni crystallographic structure, hardness, Ni−P composite phosphorus content and internal stress were significantly different when employing particles of different size. In the case of Ni−P, the presence of sub-micron size particles causes greater decrease in phosphorus content in the deposit when compared to nanoparticles. This was associated with the higher hydrogen adsorption on these particles, hence a higher restriction of the phosphorus production in this case. For composites containing sub-micron particles the internal stress was 5 times smaller than that of Ni−P electrodeposits with no reinforcement. Structurally different particles (e.g. fullerenes, nanotubes) in a single metal deposit may also exhibit different behaviour during the incorporation process [16]. Such important aspects are not extensively studied. Rather, the emphasis has been on qualitative aspects affecting the loading of particles in the deposit.

In general, the particle content in the deposit can be increased by applying appropriate agitation and the addition of metal cationic accelerants [29] plus organic surfactants [30] but also by changing the waveform of the applied current, i.e., using a pulsed current power supply [31,32].

Dispersing the reinforcing particles in the electroplating solution and obtaining a stable suspension are of crucial significance in obtaining deposits with a sufficient loading and uniform distribution of particles. Methods for improving particle dispersion include mechanical (agitation, ultrasonication [163], etc.) or chemical (altering pH value of the solution to values higher or lower than the one of particle isoelectric point) treatment of the suspension, addition of appropriate surfactants or in some cases even chemical modification of the particle surface [162].

Controlled agitation in the electrolytic bath is one of the prerequisites for achieving uniform, high quality composite coatings. In the electroplating process agitation serves many purposes: to disperse bath constituents, to keep the particles suspended in the electrolyte, to enhance the transport of particles towards the cathode surface, to increase the deposition rate, to disperse gases and overall to maintain general solution uniformity and thermal equilibrium. Agitation can be directed primarily at electrolyte mixing or it can aim at interface agitation when the goal is to reduce the thickness of the diffusion layer at the electrode/electrolyte interface. In the former case mechanical agitation and air bubbling are quite effective. Air bubbling, however, decreases solution conductivity, hence increasing the cell voltage and introducing problems of bubble formation and frothing, if surfactants are present. Although it is widely applied in industry, it must be carefully optimised to ensure a satisfactory level of internal and interfacial mixing. In the case of rotating electrodes (e.g. RDE, RCE), their movement induces both bath stirring and interface agitation, the latter resulting in the decrease of thickness of the diffusion layer [33]. These electrode geometries provide well defined fluid flow which can be used to tailor mass transport rates, hence deposit composition and morphology [34–36]. Diminishing the diffusion layer thickness enhances convective-diffusion from the bulk solution towards the electrode surface, increasing the limiting current density and the deposition rate.

In many investigations, it has been found that increased agitation enhances the loading of particles in the deposit. However, excessive agitation, resulting in turbulent flow, may lead to a lower quantity of particles in the metal deposit. This can be explained by the existence of vigorous hydrodynamic forces in the electrolyte which remove the particles from the cathode surface before they can be entrapped in the growing metal deposit [16]. Different flow regimes affect micron or nanosized particles in a different way [10,16]. Strategies for bath agitation depend on the dimensions of the electroplating cell and the scale at which the electrodeposition is being carried out. For laboratory investigations magnetic stirring, rotating disk or cylinder electrodes and parallel plate channel flow are commonly employed while in industrial processes popular methods used in open tanks include the overhead blade stirrer, the reciprocating plate plunger or a pumped recycle loop of the electrolyte [10,16]. Among all methods employed at the laboratory scale for dispersing nanoparticles, magnetic stirring is the easiest. However, it is characterized by poor reproducibility and inducing complex and eddy-prone flow. The shortcoming of this method is also the re-aggregation of nanoparticles that occurs owing to the action of van der Waal's forces once the stirring is suspended [37].

During electroplating of composite coatings, strict control of hydrodynamics in the electroplating bath is vital to obtain coatings with a high and homogeneous volume fraction of codeposited particles and its microstructure control [38,39]. This problem is especially pronounced in the large scale plating where complex hydrodynamic distribution can generate numerous problems especially in terms of the particle incorporation and the uniformity of their distribution. Many studies have been performed in order to estimate the extent of hydrodynamic influence and to mitigate concomitant problems [38,40–42]. Fransaer et al. [24] devised a particle codeposition model that took into consideration the forces affecting the particles, particularly the electrolyte fluid flow. Bozzini and colleagues [38] investigated the entrapment of particles on a rotating disc electrode with the approximation of steady laminar flow past the solid wall in a Newtonian, incompressible fluid in which rigid spheres are dispersed. Entrapment of particles and the selection of their diameter have been studied in terms of the balance of the forces acting on the particle, namely, viscous drag, Magnus and centrifugal forces, mass transport being identified as the key factor influencing the incorporation of particles.

Many solutions have been proposed, involving particular geometry and electroplating cell designs in order to achieve uniform flow conditions relative to the workpiece [3]. Gebhart and Taylor [40,41] introduced a patent in which electroplating cell was engineered so as to provide a uniform boundary layer thickness and to dampen uneven localised current distribution. Design features included directing the eductor flow in a parallel fashion across the substrate surface (Fig. 5).

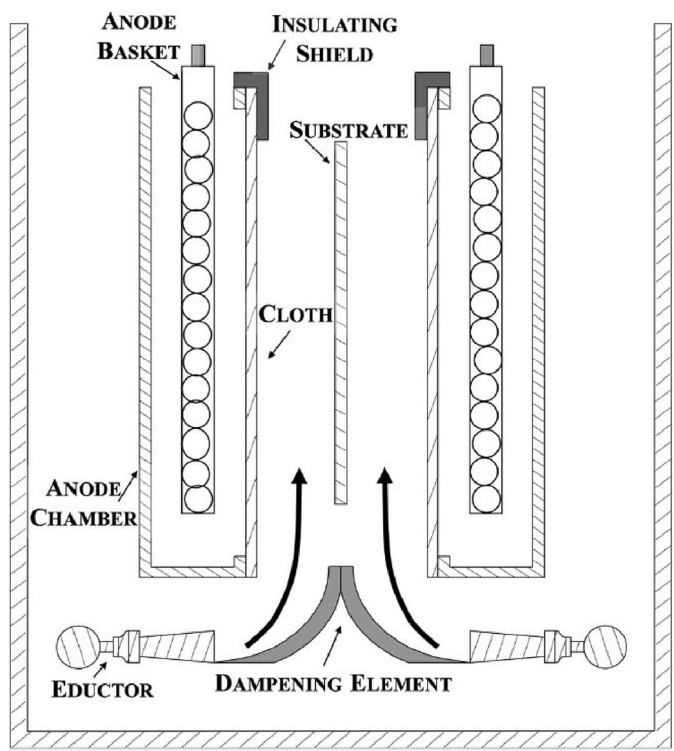

**Fig. 5.** Cross-sectional view of a cell geometry using a parallel solution flow with respect to the substrate surface. Flow is achieved by orienting the eductors below each anode chamber, directing flow to the substrate surface with the dampening element, distinct flow channels are formed between the substrate and the anode chambers.
Reprinted with permission from [41]. Copyright (2016) Electrochemical Society, Inc.

Application of an external magnetic field can help to improve the quality of fabricated composite coatings owing to a strong, localised convection arising from the magnetohydrodynamic effect which enhances the mass transfer towards the coating's surface [37]. Additionally, magnetic field enhances desorption of hydrogen and diminishes the fractional bubble coverage by reducing the mean bubble size and changing their distribution which decreases the formation of holes in the coating [43]. The magnetic field breaks down the dendrites formed on the surface of the coating and increase the nucleation resulting in a coating with finer grains. Zhou et al. [44] studied the effect of the external high parallel magnetic field on the Ni/nano-$Al_2O_3$ composites codeposition process. They found that the distribution of nanoparticles displays a network shape when applying a magnetic field and that the average diameter of single network increases with increasing the current density. A high content of nanoparticles was obtained at low current density with a magnetic field applied, while high current density was necessary to realize the same when no magnetic field was employed. Hu et al. [45] found that the application of an external magnetic field can significantly improve the mass transport process, causing an increased SiC content in the nickel matrix and modifying the obtained composite's surface morphology. Bund and coworkers [46] corroborated findings on the benefits of magnetic field application during electrodeposition. They observed an increase in the alumina particle content of the nickel electrodeposit, the effect being induced by the application of a perpendicular magnetic field.

Ultrasonic stimulation of baths containing nanoparticles can not only enhance the dispersion of particles in the bath but also improve the content and distribution of nanoparticles in the coating [163]. Cavitation bubbles which form due to instantaneous high pressure and strong shock waves, weaken the interactions between nanoparticles, resulting in an effective and even dispersion. The intensity of cavitation phenomena decreases at higher ultrasonic frequency, hence in electrodeposition lower frequencies are preferred [163]. In terms of ultrasonic power, it was observed that particle incorporation reaches a maximum at intermediate ultrasonic powers [163]. Ultrasonic cleaning baths and ultrasonic horns present the most common type of ultrasonication systems applied in the metallic matrix composite electrodeposition. Nanocomposite coatings produced under ultrasonication exhibit an enhancement in corrosion and wear resistance and improved hardness [37]. Zhu et al. [47] reported that owing to the intensity of ultrasonic agitation being much higher than that of electromagnetic stirring, the hydrogen produced concurrently with the reduction of Ni ions can be taken away from the cathode surface in time, reducing the effect of hydrogen embrittlement in the fabricated Ni matrix composites. Nishira et al. [48] studied the effects of agitation methods on the particle size distribution of PTFE aggregates in the electroless Ni–P deposition baths. Their findings indicated that deployment of an ultrasonic homogeniser is more effective than plain mechanical agitation. Similar results were obtained for the codeposition of $Al_2O_3$ particles in the Ni matrix by Garcia-Lecina et al. [165]. They obtained an enhancement of 40 nm γ-$Al_2O_3$ incorporation and its better dispersion through the application of 10 min 24 kHz, 38 W cm$^{-2}$ ultrasonic stirring before electrodeposition. Zanella et al. [166] investigated the influence of applying ultrasonication during the electrodeposition process of Ni-nano SiC composites under DC and PC conditions. Deposits with improved microhardness and corrosion resistance were obtained with marked synergic effect of pulse plating and ultrasonication on final properties. Chou and coworkers [59] applied 20 min ultrasonic agitation to disperse 0.3 μm SiC particles before Ni–P composite electrodeposition. Aggregation of SiC particles has not been completely prevented; however, they obtained composites with lower residual stress compared with pure Ni–P alloy fabricated at different current densities.

Use of appropriate surfactants to ensure a stable dispersion of particles in the electroplating solution is a critical factor in composite electrodeposition. Colloidal and nanosuspensions are thermodynamically unstable, due to the large surface energy of the particles, hence preventing agglomeration via agitation and addition of surfactants is vital. The stabilisation of particles in a high ionic strength medium, such as an electroplating bath, is particularly challenging. In the absence of suitable additives quality of the dispersion of particles tends to be low and the influence of purely physical parameters such as the geometry and the position of the workpiece becomes more important [10]. Cationic (e.g. cetyltrimethylammonium bromide), anionic (e.g. sodium dodecyl sulfate) and nonionic (e.g. octylphenol ethoxylate, polyethylene glycol) surfactants, plus their combinations, can be used to control surface charge of particles. High values of zeta potential of like sign maximize electrostatic repulsion between the particles, minimising their aggregation and helping to ensure a stable particle suspension. An excessive surfactant concentration in the bath should to be avoided, however, as it can modify electrocrystallisation, leading to enhanced internal stress and reduced strength in the deposit [49]. In the case of Ni–P composites, common surfactants are: SDS [69,70,97] and CTAB [14,132,68]. Besides electrostatic stabilisation, colloids can be stabilised through adsorption or chemical attachment of polymeric molecules (steric stabilisation). This type of nanoparticle suspension stabilisation mechanism has not been reported in the case of Ni–P composite electrodeposition.

Most published studies which employ techniques for coatings deposition are applied only at the laboratory sale. Up-scaling of these processes, especially in the case of more complicated compositions and structures of coatings (nanostructured, gradient, nanolattice films), larger pieces or more complicated shapes presents a huge technical and economical challenge. This is due to the introduction of many new variables which induce many modifications to the process but also owing to the difficulties related to maintaining particles in the dispersed state when using large volumes [50,51]. Deposit properties can change with scale and process conditions must be well defined and carefully optimised. More attention needs to be paid to technology transfer and feasibility trials on an industrial level.

It is common practice, especially in industrial applications, to use the same electrolytic bath for fabrication of many deposits over extended times. Bath components become depleted or enriched, the composition continuously changing. It is vital to keep track of the extent and nature of the bath composition evolving, electrolyte degradation and ageing under the specific deposition conditions to secure deposits of uniform and consistent quality. Very few studies have been devoted to the importance of Ni–P electrolytic bath ageing, despite the relatively high costs involved in bath maintenance and replacement.

## 3. Ni–P electrodeposits reinforced by ceramic particles

Mitigating problems related to corrosion and wear of materials is the subject of major focus in the modern world. In the 1970s, Rabinowicz conducted a large study of US-based industry in order to determine the key causes of premature machine failures. In his "loss of usefulness" study [52] he later pointed out that 50% of all failures of machine parts are being caused by wear and can be traced back to inadequate lubrication strategies and 20% by corrosion which in total can account for up to 6–7% of GNP. It is estimated that the world's economy loses billions of dollars every year due to energy losses caused by machinery depreciation and eventual failure. In the light of these facts it is easy to appreciate the immense significance of exploring and adopting the most effective surface engineering practices.

Hard and wear resistant coatings are commonly deposited on tools which are used for severe cutting, forming and casting applications, where the conditions typically result in high temperatures, increased mechanical loads and pronounced wear [53]. Hard Cr coatings are renowned for their exceptional wear and corrosion resistance but also materials such as nitrides, carbides, oxides, borides and carbon-based ones are extensively used in hard coatings technology owing to their outstanding mechanical and tribological properties [54]. Excellent chemical stability and oxidation resistance in severe environments are also their recurring features.

Functional properties are directly related to deposit structure. However, multi-component materials containing ceramic inclusions can provide improved deposit properties owing to their chemical and mechanical properties together with induced structural and compositional modifications. This is heavily exploited in the deposition of Ni–P composites containing different reinforcement phases and several examples will be given.

### 3.1. Ni–P electrodeposits reinforced by SiC

The most commonly studied Ni–P based composites produced by electrodeposition contain SiC particles. SiC is employed in a wide range of industrial applications owing to its superior physical and chemical properties, including excellent room and high temperature hardness, wear resistance and chemical durability [55]. Many findings corroborate that the addition of homogeneously dispersed SiC nano or micron sized particles to the Ni–P alloy matrix leads to achievement of tough, dispersion hardened coatings with improved properties [56–58]. Additionally, employing this type of reinforcement results in the most sensible balance between the amelioration of performance characteristics and the process cost effectiveness.

The majority of studies demonstrate that increasing the SiC particles concentration in the electrolytic bath results in the increase of the SiC content and decrease of the phosphorus content in the fabricated Ni–P based composites [31,32,59]. Latter was attributed to the restriction of phosphorus production owing to the enhanced adsorption of $H^+$ ions on the surface of SiC particles [31]. Increasing particle bath load is beneficial in terms of increasing SiC content in the deposit, up to a certain level. An excessive SiC loading in the electrolyte leads to saturation, particle agglomeration and subsequent decrease of SiC content in the coating [60]. Yuan et al. [57] studied the codeposition of 50 nm SiC particles within the Ni–P matrix. An increase of SiC concentration in the bath to $10\,g\,L^{-1}$ resulted in significant agglomeration.

Smaller particles tend to be more difficult to codeposit [61]. Nanosized particles exhibit higher tendency towards agglomeration in the electroplating solution which can be detrimental in terms of obtained deposit properties. Garcia et al. [62] showed a higher particle volume fraction in the case of 5 μm SiC particles, compared to 0.7 or 0.3 μm particles. A maximum content of 23 vol% 5 μm SiC was achieved at a 100 g/l bath loading of particles. Similarly, Wang et al. [63] demonstrated that the ultrafine SiC was more difficult to co-deposit than the coarse SiC, and that the rate determining step was controlled by a transition from weak to strong adsorption (in accordance with the Guglielmi model). Lee et al. [64] demonstrated that the nanosized SiC particles are more difficult to codeposit in the Ni matrix than micron sized particles. Furthermore, they observed that the codeposition of α-SiC particles is easier than β-SiC owing to the more negative zeta potential of α-SiC. Powders with high zeta potential (absolute value) usually exhibit higher stability in slurries than powders with low zeta potential [65]. Accordingly, the stability of slurries can be investigated by zeta potential measurement [66]. As a general rule, absolute zeta potential values above 30 mV provide charge stabilisation of nano particle suspension [161]. Malfatti and coworkers [67] investigated the codeposition of SiC particles with large granulometric distribution within the Ni−P matrix. They found however that even though volumetric fraction of incorporated particles reaches a plateau when increasing particle load, the incorporated particle number per unit area continues to increase accompanied by the reduction in the incorporated particle size, demonstrating the selective nature of the process (Fig. 6). According to these authors, this was due to a higher probability of mechanical interactions between larger particles and their subsequent ejection, which reduces the likelihood of large particles incorporation in the Ni−P deposit. This finding highlights the significance of taking into the account both volume fraction and number of particles per unit area.

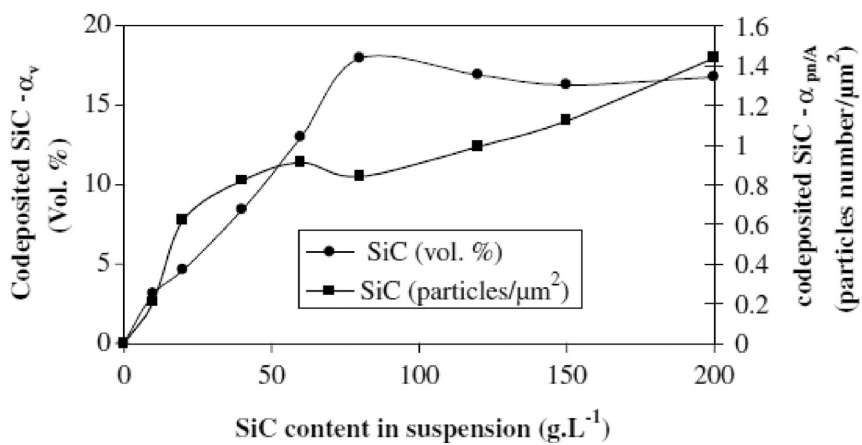

Fig. 6. Amount of SiC particles in NiP/SiC$_{600}$ composite coatings as a function of the particle concentration in bath. Reprinted with permission from [67]. Copyright (2005) Elsevier.

Application of suitable additives is very important when dealing with SiC particles electrolytic codeposition. Narasimman et al. [30] studied the effect of the presence of several different surfactants (TX, SDS, SAC, CTAB, TMAI, TMAH) on the obtained volume fraction of incorporated β-SiC particles in the Ni matrix. Their findings indicated that TMAH performs the best in terms of obtaining deposits with highest volume percentage of incorporated particles and their most homogeneous distribution. In the case of Ni−P matrix, Hou et al. [68] found that the addition of surfactant CTAB can decrease the agglomeration of inherently hydrophobic SiC particles in the aqueous solution. They observed a trend of increasing sub-micron SiC particles content in the deposited Ni layer with increasing concentration of the surfactant in the electrolyte. 11.5 vol% of incorporated SiC was obtained at a $20\,g\,L^{-1}$ particle load in the bath in the presence of $0.084\,mol\,L^{-1}$ CTAB.

Malfatti et al. [69] investigated co-electrodeposition of SiC particles possessing a large granulometric distribution (average diameter 600 nm) within the Ni−P matrix. They found that the addition of anionic (e.g., SDS) and cationic (e.g., CTAHS) surfactants decreases the amount of codeposited SiC particles coupled with the increase in size of the embedded particles. Obtained results suggested that the particle incorporation was practically independent of the surface charge on the surfactant.

Saccharin is a commonly used grain refiner in the Ni−P alloy electroplating, although it is seldom appreciated that this additive is multifunctional, also acting as a surfactant, stress modifier and leveller. Addition of saccharin significantly improves cathode current efficiency of Ni−P electrodeposition [61,70] and decreases the internal stress of the deposit [71]. This compound however can lead to composite coatings having lower contents of SiC particles and of phosphorus, compared to deposits from saccharin-free baths [72].

With suitable control of operating conditions, codeposition of SiC particles within a Ni−P matrix results in an increased hardness, improved corrosion and wear resistance and a lower coefficient of friction of the alloy against steel [56,57,73]. Reduced residual stress and eliminated surface cracking have also been reported, together with increased ductility [32,59,74]. Ni-P/SiC composites, however, can exhibit higher surface roughness than pure Ni−P alloy electrodeposits [75].

As in the case of pure Ni−P electrodeposits, owing to precipitation hardening, annealing improves hardness of Ni-P/SiC composites substantially. Martínez-Hernández et al. [73] studied the incorporation of 100 nm SiC particles in the Ni−P matrix via electrodeposition. The Ni-P/SiC composites (approx. 0.5 at. % SiC) exhibited higher hardness (600 HV) when compared to pure Ni−P alloy fabricated under the same electroplating conditions (430 HV). After annealing at 500 °C deposits attained very high hardness (1453 HV), which was greater than the hardness of a typical hard Cr coating. With further thermal treatment above 500 °C hardness decreased sharply. Annealing led to a significant decrease of SiC content in the coating. This was owing to the particle detachment from the upper and middle layers of the coating during the thermal treatment. Vaillant et al. [76] observed that Ni-P/SiC electrodeposits containing > 15 at.% of P upon heat treatment at 420 °C exhibited hardness (1100 HV) and mass loss similar to those obtained for hard Cr.

Wang and coworkers [56] investigated the effect of post-deposition heat treatment on structural, mechanical and tribological properties of Ni-P/SiC nanocomposite coatings. SiC particles with a bath concentration of $5\,g\,L^{-1}$ and an average size in the range 45–55 nm were employed. Obtained results indicated the formation of $Ni_3P$ phase in the annealed coatings regardless of the treatment temperature. Gradual augmentation of deposits hardness was observed upon annealing at temperatures up to 400 °C while further increase of annealing temperature caused a reduction in hardness. Nevertheless, all annealed coatings were harder than as-plated deposits. Improved tribological properties were however observed for composite coatings annealed at temperatures ≥ 400 °C. This was associated with the tribochemical reaction of $P_2O_5$ oxide with water from the environment and the formation of $H_3PO_4$ which tends to exhibit a lubricating effect. In spite of the lower hardness compared to other coatings, Ni-P/SiC composite annealed at 500 °C exhibited the lowest friction coefficient (0.51) and wear rate ($7.8 \times 10^{-6}\,mm^3\,N^{-1}\,m^{-1}$) due to sufficient formation of $H_3PO_4$ and improved surface morphology. Ni-P/SiC deposits annealed at 350 °C exhibiting the highest hardness of approx. 830 HV demonstrated the worst tribological properties, due to their rough surface morphology and scarcity of the solid lubricant.

Martinez-Hernandez et al. [73] detected a decrease in the wear volume of Ni−P matrix upon addition of 100 nm SiC particles and a very low coefficient of friction comparable to the one of hard Cr coatings. In the case of Ni, Cheng et al. [77] established that the weight loss of friction of Ni/SiC composite coatings decreases with increasing 5 μm SiC content in the coating, exhibiting a minimum at the optimum value of second phase particle content (20 wt%). Further increase in SiC content beyond the optimum amount caused a deterioration in tribological properties. This effect was attributed to the excessive quantity of SiC particles that formed a soft constituent. Garcia et al. [62] reported that a decrease in size of SiC particles improves the wear resistance of the Ni/SiC composite coatings owing to the resultant particle number density increase in the deposit rather than their volume fraction augmentation. Hou et al. [68] however stated subsequently that the assumption used in the previously mentioned work of SiC particles being mono-disperse and spherical can cause significant discrepancies and oversimplifies the codeposition process interpretation as it ignores the possibility of particle agglomeration in the plating bath which would influence the vol% of SiC in the deposition layer. They employed surfactant CTAB to prevent the agglomeration of SiC sub-micron particles and found that the wear resistance of Ni/SiC electrodeposit increases with increasing SiC vol% in the deposit layer. Aslanyan et al. [75] observed in the case of Ni−P matrix, that the wear rate exhibits a growth when the SiC content increases in the coatings. They detected a lower wear rate for annealed pure Ni−P coatings when compared to composites, even though composites exhibit higher hardness [58,75,78]. This was justified with the increased sensitivity towards crack formation around SiC particles in the case of composites. Aslanyan and colleagues [58,75,79] found that, in spite of the admixtures of SiC and the heat treatment, the wear of the electrolytic Ni−P coatings is by abrasion and oxidation. Wear properties of Ni-P/SiC composite coatings were shown to depend on the type of the sliding test. In the case of the unidirectional sliding test abrasive wear was noticed, whereas in the bi-directional sliding test oxidational wear was revealed [79].

Corrosion resistance is another crucial aspect when it comes to the application of the fabricated coatings. Yuan et al. [57] observed that the addition of nano-SiC particles improves the corrosion resistance of the Ni−P electrodeposits. This was justified by the assumption of the SiC particles presence decreasing the exposed area of the deposit and increasing the density of boundaries between the nanoparticles and the matrix. The best corrosion resistance in 3.5 wt% NaCl solution was obtained for the deposits having an intermediate SiC content (3.07 wt %). Further increase of SiC amount caused corrosion resistance deterioration owing to particle agglomerates occurrence which led to uneven surface and hence increased corrosion susceptibility. However, all composites exhibited better corrosion resistance when compared to as-plated amorphous Ni–P deposits.

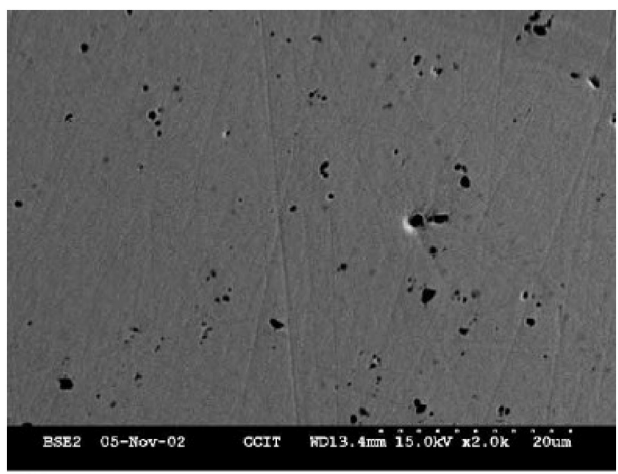

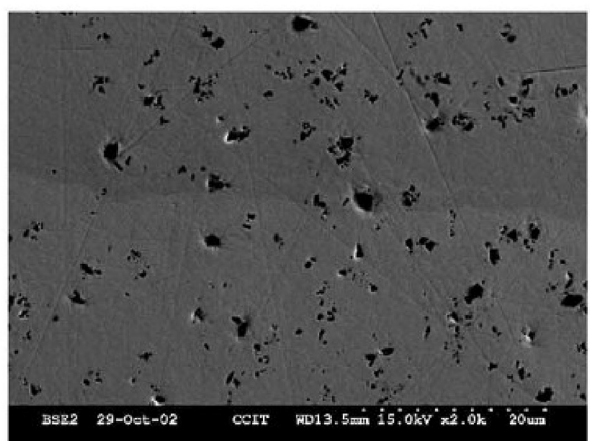

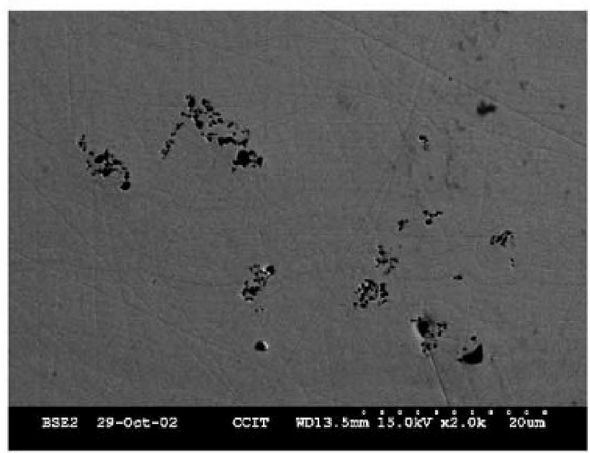

**Fig. 7.** SEM micrographs of the polished cross-section of Ni-P-SiC deposits at various current densities and SiC concentrations: (a) 20 A dm$^{-2}$, SiC 1 g L$^{-1}$; (b) 20 A dm$^{-2}$, SiC 10 g L$^{-1}$; (c) 5 A dm$^{-2}$, SiC 10 g L$^{-1}$.
Reprinted with permission from [59]. Copyright (2005) Elsevier.

Malfatti and coworkers [67] observed also that the presence of SiC particles in the Ni–P matrix improves its electrochemical behaviour in 0.6 mol L$^{-1}$ NaCl solution. However, they found that, in the case of electrodeposits possessing the same volumetric fraction of SiC particles, the ones containing particles of smaller size exhibit higher corrosion current densities, which was attributed to the increased matrix/particle interface area. Further deterioration of corrosion resistance for composites was observed upon thermal treatment due to concurrent shrinkage of the metallic matrix.

Some findings indicate that an increase in current density influences the codeposition of SiC particles within Ni–P matrix in a negative way [4,61]. However other works demonstrate positive correlation between the two quantities [67,59]. Fig. 7(b) and (c) show the increase of SiC content in the deposit with increasing the current density [59].

Employing pulse plating instead of direct current deposition at the same average current density leads to composite coatings with higher incorporation rate of SiC particles and a more uniform particle distribution [31,32], the effect being more significant at higher SiC concentrations in the bath [32]. Pulse current electrodeposition results in deposits possessing overall better properties: reduced porosity, surface roughness, improved ductility, hardness, wear resistance. This is owing to the numerous beneficial effects of this plating regime: replenishing of the depleted diffusion layer during the time when the current is suspended, recrystallization of thermodynamically unstable small grains, hydrogen bubbles desorption and many others [80]. Spyrellis et al. [81] studied the codeposition of 1 μm SiC particles in the Ni–P matrix. They observed that in the PC regime the amount of co-deposited SiC particles (5–8 wt%) is twice as high as the one obtained by DC plating. Low duty cycle values and high frequencies (10% and 100 Hz) tend to favor embedment of SiC particles according to the authors of [31], while others find low frequencies in combination with low duty cycles to be beneficial [61].

Zoikis-Karathanasis and coworkers [31] observed that in the electrodeposition of 1 μm SiC particles from an additive-free bath pulse plating leads to a higher percentage of the codeposited SiC particles regardless of the rotation velocity of the employed RDE. SiC content exhibited a maximum value at the optimum rotation velocity which increases the convective flow towards the disc electrode. The highest percentage of incorporated SiC particles (for 20 g L$^{-1}$ particle load) was observed for the coating produced under duty cycle 10%, frequency 100 Hz and 700 rpm RDE, reaching approx. 22 wt%. Annealing of the obtained composite coatings resulted in the improvement of mechanical properties. Deposits with low phosphorus content exhibited XRD patterns containing main diffraction peaks of Ni, those with intermediate phosphorus content revealed peaks of Ni$_3$P (Fig. 8), while deposits with phosphorus content higher than 11.5 wt% demonstrated the presence of an additional Ni$_2$P phase (Fig. 9). Under certain conditions of pulse plating, as-plated deposits contained a Ni$_{12}$P$_5$ crystalline phase which persisted after annealing.

Hansal et al. [61] investigated the influence of pulse reverse plating on the incorporation of micron and sub-micron SiC particles in the Ni–P matrix. Their results demonstrated that the particle incorporation rate decreases with increasing current density and shows no significant dependency from the pH value in the investigated range (pH 1–2). A decrease in frequency below 5 Hz with a duty cycle of 20% resulted in a higher particle incorporation rate (150 g L$^{-1}$ bath loading) reaching approx. 3 wt%. Although incorporation of the particles seemed to be hindered at short pulse times, higher frequencies were beneficial in obtaining deposits with optimal hardness and wear resistance owing to low phosphorus content and intermediate particle amount. Applying unipolar pulses led to improved hardness and wear resistance, while bipolar pulses generated higher amount of phosphorus in the deposit causing a deterioration in mechanical properties and cathode current efficiency. Columnar growth was observed in the case of unipolar deposition, while the application of bipolar pulses induced a change to a lamellar structure.

According to several studies [31,61], deposits obtained under PC conditions exhibit higher microhardness owing to the higher SiC content when compared to those obtained by DC deposition. According to others [32], hardness of composites produced under PC regime is lower even though SiC codeposition rate is higher. Phosphorus content in the coating has a secondary effect on hardness according to the authors of [31]. Others find that phosphorus amount in the deposit is the main factor influencing hardness while SiC exhibits an indirect effect by influencing the composite microstructure [59]. The latter applies in the case of deposits with low SiC incorporation rate. When the SiC concentration in the bath is low the fall in microhardness with the increase of SiC content in the deposit is caused by the decrease of phosphorus content in the deposit. At higher bath loadings the effect of the increase of SiC quantity in the coating becomes more significant than the effect of phosphorus content decrease and the microhardness increases [32]. In general, improved hardness (ca. 700 HV) upon addition of SiC particles to Ni–P matrix is observed [31,32,59,61] with reports of hardness values which exceed 1000 HV after thermal treatment [73,76].

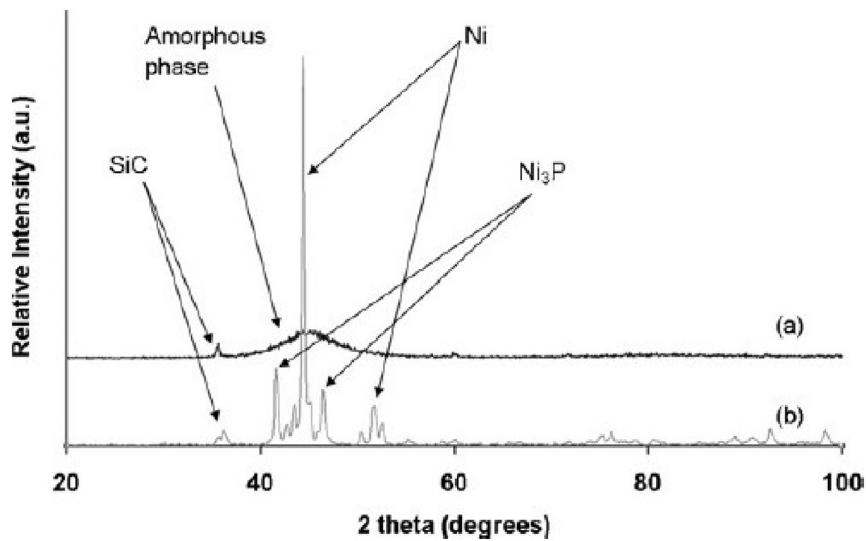

**Fig. 8.** XRD patterns (a) before and (b) after thermal treatment of a Ni-P/SiC deposit with a P content of 4.35 wt% and a SiC content of 9.6 wt%, produced under PC conditions (10% duty cycle and 0.1 Hz frequency).
Reprinted with permission from [31]. Copyright (2010) Elsevier.

### 3.2. Ni–P electrodeposits reinforced by $B_4C$

Fields of interest for Ni-P/$B_4$C composite coatings include those in which highly corrosion and wear resistant deposits are required. $B_4$C possesses high hardness (38 GPa Vickers), elastic modulus (460 GPa) and fracture toughness (3.5 MPa m$^{-2}$), low specific gravity and increased neutron absorption.

Bozzini et al. [82] investigated corrosion and erosion-corrosion behaviour of fabricated Ni-P/$B_4$C electrodeposits. The presence of 8 vol% 70 μm $B_4$C particles was correlated with higher crystallinity of the matrix at 5% phosphorus content, increase in microhardness (approx. 615 HV), a more noble breakdown potential and a lower current density plateau for the composites. In general composites displayed improved corrosion and erosion-corrosion resistance in aerated, slightly acidic, chloride solutions. A positive correlation between $B_4$C volume fraction and current density was reported.

Bernasconi et al. [83] studied the codeposition of 2 μm $B_4$C particles within the Ni–P matrix. Two compositions of Ni–P, one with a low phosphorus content in the order of 4 wt% and another with a high phosphorus content of about 12 wt%, were deposited from a modified Watts nickel electrolyte using both direct and pulse current plating regimes. PC plating was found to induce a significant augmentation of both phosphorus content and the quantity of co-deposited particles.

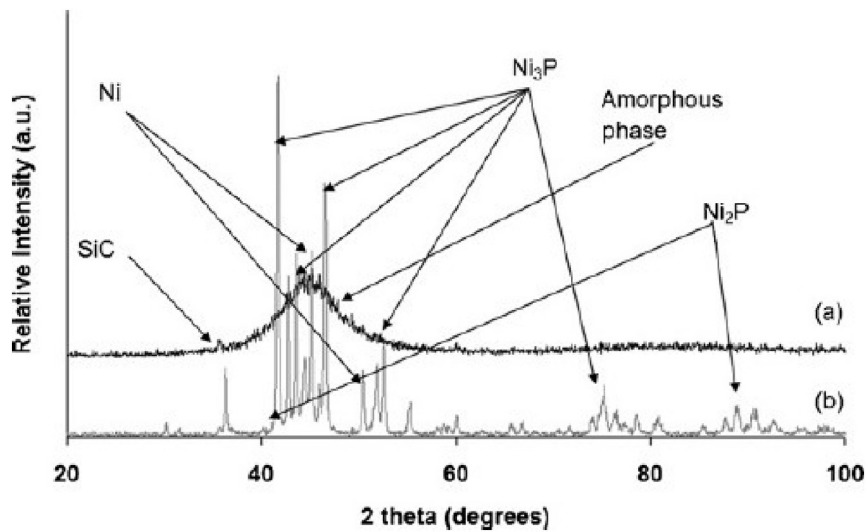

**Fig. 9.** XRD patterns (a) before and (b) after thermal treatment of a Ni-P/SiC deposit with P and SiC contents of 11.6 and 1.2 wt%, respectively, produced under PC conditions (90% duty cycle and 1 Hz frequency).
Reprinted with permission from [31]. Copyright (2010) Elsevier.

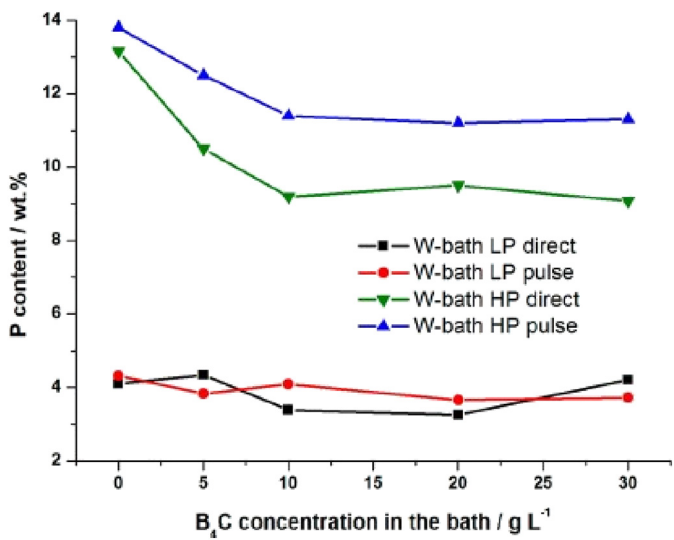

**Fig. 10.** Phosphorus content as a function of particle concentration in the bath and the type of power control for electrolysis, DC or PC.
Reprinted with permission from [83]. Copyright (2017) Taylor & Francis.

Maximum particle content was obtained in the composite plated from low phosphorus bath. It was estimated that 25% of the deposit area was occupied by particles. Phosphorus content exhibited a decreasing tendency as particle content increased, the effect being more evident when the high phosphorus formulation was considered (Fig. 10). The hardness reached a maximum of about 830 HV in the case of PC plating from a low phosphorus bath containing $5\,\mathrm{g\,L^{-1}}$ of reinforcing particles. PC control resulted in improved mechanical properties and a more uniform particle distribution in the deposit.

### 3.3. Ni–P electrodeposits reinforced by WC

WC is an extremely hard ceramic material (> 2000 HV) that possesses high thermal and chemical stability. WC maintains its hardness value up to 1400 °C, possesses wear resistance which is better than for tool steel and low electrical resistivity of ca. 0.2 μΩ m.

Spyrellis and colleagues [81] investigated the effects of electrolysis parameters on the structure and morphology of Ni and Ni–P matrix composite coatings reinforced by 200 nm WC particles in DC and PC electrodeposition conditions. Ni-P/WC coatings were characterized by the highest phosphorus content when compared to pure Ni–P and Ni-P/SiC composite coatings. Employing PC plating led to an increase of the percentage of incorporated particles. 30 wt% of WC particles was achieved, while in the case of Ni-P/SiC composites percentage of embedded SiC particles was 5–8 wt%. In both cases percentages of co-deposited reinforcing particles obtained with PC plating were more than twice higher than those achieved by DC plating. Additionally, embedding of ceramic particles modified in various ways Ni electrocrystallization process, while Ni–P amorphous matrix was not affected by the occlusion of the particles having also greater amount of non-agglomerated particles of the solid phase that were embedded as compared to the polycrystalline Ni matrix. Hardness of as-plated composites was approx. 6.9 GPa which presented a significant improvement when compared to the hardness of pure Ni–P deposits (approx. 4.8 GPa) obtained under the same conditions. Annealing of PC plated Ni-P/WC composite coatings at 400 °C resulted in complete crystallization, revealing Ni, Ni$_2$P and Ni$_3$P phases, accompanied by a significant increase in microhardness [84]. Maximum value of hardness obtained after heat treatment was 18.6 ± 1 GPa.

Xu and colleagues [85] performed electro-brush plating of Ni-P/WC composite coatings. The deposits were reinforced by the electric contact strengthening (ECS). ECS is a surface treatment which is based on electric contact resistive heating localised at the surface of the work piece, with the aim of improving the coating properties and bonding strength. Results obtained indicated that the application of ECS reduces the number of defects and results in a denser microstructure. Hardness and wear resistance of the coatings were also improved.

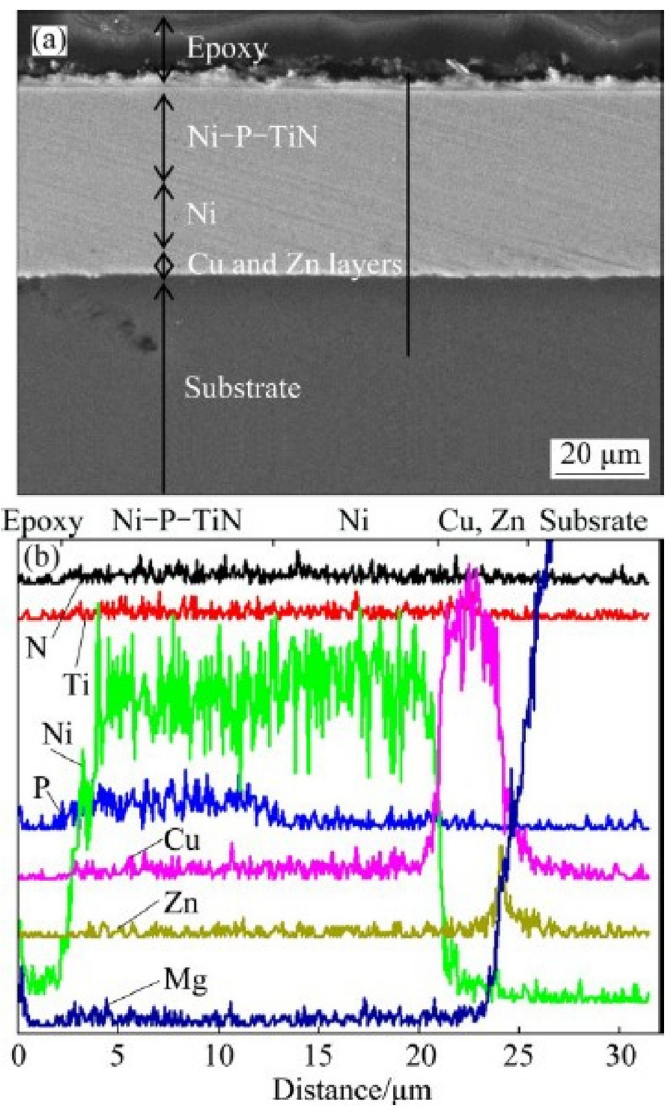

**Fig. 11.** (a) Morphology of a cross section and (b) an EDS line profile for a Ni-P/TiN coating electrodeposited at a current density of $8\,\mathrm{A\,dm^{-2}}$ at pH 2.0.
Reprinted with permission from [86]. Copyright (2016) Elsevier.

### 3.4. Ni–P electrodeposits reinforced by TiN

Zhou et al. [86] performed the co-electrodeposition of TiN nano particles within the Ni–P metal matrix. Fabricated Ni-P/TiN composites demonstrated good adhesion on the magnesium substrate owing to the series of applied pretreatments. Fig. 11 shows the cross-section morphology of the obtained deposit. Sequence of layers enables good adhesion and fabrication of a compact and pore free Ni-P/TiN electrodeposit. Hardness of the composite was higher when compared to the hardness of pure Ni–P alloy. Upon annealing at 400 °C for 1 h, the hardness of the Ni–P alloy and Ni-P/TiN composite was 653 HV and 855 HV, respectively. Electrodeposition of the coating provided a significant improvement in the long term corrosion resistance of a magnesium substrate in a 3.5% NaCl solution. Pure Ni–P electrodeposits performed better in terms of short-term corrosion protection owing to the phosphorous rich layer formation, however in the long term the presence of TiN nano-articles contributed to the inhibition of pitting attack penetration and brought more lasting corrosion stability of the substrate. The corrosion potential during immersion of the composite in a 3.5% NaCl solution was stable at approximately −0.4 V vs. SCE after > 1600 h, in contrast to a Ni−P coating that demonstrated significant decrease of corrosion resistance after 170 h.

Ma et al. [87] fabricated Ni-P/TiN nanocomposites on steel sheets through magnetic electrodeposition technique. Magnetic field application resulted in improved microstructure and decreased grain size owing to the magnetohydrodynamic effect that reduced the viscosity of the plating bath and increased movement rates of $Ni^{2+}$ ions and TiN particles which promoted the codeposition. Thermal treatment brought a significant enhancement of mechanical and electro-chemical properties. Composites heat treated at 500 °C for 10 min demonstrated the highest microhardness (approx. 919 HV) and best corrosion resistance in a 0.55 mol $L^{-1}$ NaCl solution.

### 3.5. Ni−P electrodeposits reinforced by $TiO_2$

Titanium oxides are used as activating agents for cathode materials in numerous electrochemical processes [88]. In most studies, Ni-P/$TiO_2$ composites fabrication was achieved through electroless deposition. Lee et al. [89] established that in autocatalytic deposition addition of $TiO_2$ nanoparticles slightly decreases the phosphorus content of the deposit.

Losiewicz and coworkers [90] asserted that the introduction of $TiO_2$ to amorphous Ni−P through electrodeposition results in the increase of the rate of the hydrogen evolution in comparison with conventional Ni−P layers in both acid and alkaline environments. The authors attributed this effect to the presence of $TiO_2$, its protonated states and/or redox forms which exert an electrocatalytic effect on the hydrogen evolution reaction (HER) and an increase in the real surface area.

Gierlotka et al. [91] found that the heat treatment does not bear an appreciable influence on the rate of hydrogen evolution of a Ni−P electrode however induces a slight inhibition of hydrogen evolution of a Ni-P/$TiO_2$ electrode in the acidic environment. Heat treatment was found to be beneficial by the authors of [92]. They observed the increase in catalytic activity occurring after heating of Ni-P + $TiO_2$ + Ti layers which was attributed to $TiO_2$ reduction and formation of non-stoichiometric Ti oxides.

Losiewicz and colleagues [93] made an attempt to elaborate the relationship between the obtained particle content in the electrodeposited Ni-P/$TiO_2$ composite coatings and the key electrodeposition parameters, such as cathodic current density, bath temperature as well as the content of $TiO_2$ powder suspended in the electrolytic bath. They employed the Hartley's polyselective quasi D optimum plan of experiments and found within the limits of their study that the porous composite Ni-P/$TiO_2$ coating with the maximal $TiO_2$ content of about 28.7 at.% can be obtained from the suspension bath containing 99 g $L^{-1}$ of $TiO_2$ at 40 °C using the cathodic current density of 5 A $dm^{-2}$.

Sadeghi [2] found $TiO_2$ nanoparticles incorporation (1.8 at. %) in the Ni−P electrodeposits to be greater when compared to the incorporation of SiC (0.4 at. %) and $Al_2O_3$ (1.6 at. %) particles having the same nominal size (ca. 50 nm) and bath loading (20 g $L^{-1}$), electrodeposition being carried out under the same working conditions. This type of reinforcement led to the highest decline in the internal stress of the deposit.

### 3.6. Ni−P electrodeposits reinforced by $SiO_2$

$SiO_2$ particles are mostly incorporated in the Ni−P metallic matrix by the means of electroless deposition [89,94–98]. Codeposition of $SiO_2$ particles from an aqueous solution into the metal matrix is limited, due the inherent hydrophilicity of the $SiO_2$ particles which makes the codeposition of $SiO_2$ into metal matrix from an aqueous solvent extremely difficult. It is reported that the maximum content of micron-sized $SiO_2$ particles in Ni matrix composite coating obtained from aqueous electrolytes is < 6 wt% and that the content of co-deposited particles decreases to 1 wt% or less (without additives) when the particle size changes from micrometer to nanometer dimensions [99]. Nowak and colleagues [100] made an attempt to elaborate why the codeposition of $SiO_2$ particles is more difficult when compared to SiC. Electrochemical impedance spectroscopy measurements demonstrated that the presence of $SiO_2$ increases while the presence of SiC decreases the electrode capacitance. Positive influence of the $SiO_2$ particles on the cathode surface roughness, hence increase of the surface area, was the only effect detected for this type of particles while SiC particles also block the cathode surface demonstrating the balance effect of the two phenomena.

As an alternative to classical aqueous and organic electrolytes Li and colleagues [99] employed a new type of ionic liquid: a deep eutectic solvent (DES) as an electrolyte in order to produce Ni matrix composite coatings containing $SiO_2$ nanoparticles (15–30 nm) by employing pulse current electrodeposition. Effective and stable dispersion of $SiO_2$ nanoparticles in choline chloride/ethylene glycol-based DES was achieved without any stabilizing additives due to its higher viscosity and ionic strength when compared to aqueous solutions. Incorporation of $SiO_2$ nanoparticles in this environment almost reached the level of incorporated $SiO_2$ micron size particles from an aqueous electrolyte (4.69 wt%). Authors determined that the presence of $SiO_2$ particles bears influence on the nucleation/growth process, microstructure and composition of Ni coating. Ni/$SiO_2$ composite coatings exhibited much better corrosion resistance compared to pure Ni, which increased with increasing $SiO_2$ content in the coatings.

Research studies on electrodeposition of these particles in a Ni−P matrix are sparse. Xu and colleagues [101] co-deposited nanostructured $SiO_2$ and $CeO_2$ particles in the Ni-W-P based matrix. The maximum microhardness of Ni-W-P-$SiO_2$-$CeO_2$ composite coatings was obtained after heat treatment at 400 °C (1338 HV) and it was higher for composites containing both kinds of particles when compared to composites containing each individual type of reinforcement. Abrasion resistance of the composites also presented a significant improvement. Maximum abrasion resistance was achieved for Ni-W-P-$SiO_2$-$CeO_2$ composites after heat treatment (0.78 mg $cm^{-2}$ $h^{-1}$). These deposits exhibited amorphous structure in the as-plated state and a transition to crystalline state after thermal treatment. More favourable high temperature oxidation resistance of these films was also established [102].

### 3.7. Ni−P electrodeposits reinforced by $Al_2O_3$

$Al_2O_3$ particles possess superior properties such as moderate costs, chemical stability, good microhardness and wear resistance [6]. $Al_2O_3$ nanoparticles are however very arduous to disperse [103]. Work by Kuo et al. [104] demonstrated that applying ultrasonic oscillation to the electrolytic bath could more efficiently improve the dispersion of nano $Al_2O_3$ than applying a surfactant. However, a higher volume content of $Al_2O_3$ (6.8 vol%) particles in the deposit is obtained when employing a cationic surfactant CTAB, owing to its chemical adsorption on the particles and resultant increased attractive force between the positively charged agglomerates and the cathode.

Composites of Ni−P containing $Al_2O_3$ particles are mostly produced by electroless deposition technique. Sheu and colleagues [6] investigated the influence of employing PC plating for the codeposition of $Al_2O_3$ nanoparticles in the Ni−P matrix without the presence of surfactants. They concluded that the percentage of $Al_2O_3$ in the deposit increases with increasing its content in the electrolytic bath (Fig. 12) and with increasing the current density, indicating that process is adsorption controlled in the used current density range. PC plating led to a higher percentage of embedded particles and their more uniform distribution in the matrix when compared to DC plating. High duty cycles (0.7), corresponding to smaller peak current densities, and high frequencies (1000 Hz) were found to be beneficial when it comes to the particle incorporation percentage and composites' microhardness.

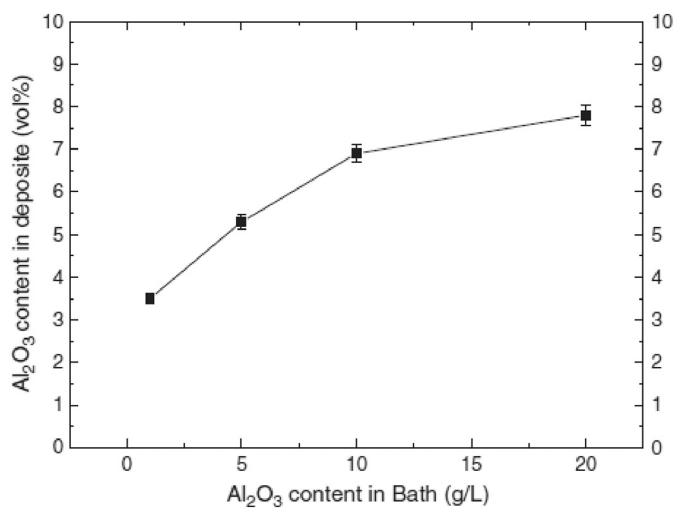

**Fig. 12.** Volume percentage of codeposited $Al_2O_3$ nanoparticles in the coatings prepared by direct plating at a current density of 5 A dm$^{-2}$ as a function of $Al_2O_3$ concentration in the plating bath.
Reprinted with permission from [6]. Copyright (2013) Elsevier.

Maximum amount of $Al_2O_3$ was found to be 24.6 vol% and maximum microhardness of the as-plated deposits was 690.3 HV. The increase in particle incorporation with decreasing peak current density they asserted is in line with the model proposed by Celis et al. [23]. They stated that the increase in peak current density leads to preferential reduction of metal ions at the cathode owing to their higher mobility, compared to particles having an adsorbed sheath of ions.

Sadeghi [2] performed the codeposition of $Al_2O_3$ particles of nano and sub-micron size within the Ni–P metallic matrix. An increased codeposition of sub-micron $Al_2O_3$ (~7 at.% Al) was achieved when compared to nanosized particles (ca. 1 at.% Al). He observed the reduction in phosphorous codeposition and crack formation upon addition of the reinforcement particles. Phosphorus content in the pure Ni–P alloy was 5.8 at.% compared to 3.5 at.% for the composite containing sub-micron $Al_2O_3$ particles and 4.5 at.% for a composite containing nano-$Al_2O_3$. Addition of both types of particle induced a lowering of internal stress in the deposit. Internal stress reduction was greater in the case of sub-micron size $Al_2O_3$. A slightly lower hardness was observed for composites (ca. 550 HV) compared to a pure Ni–P alloy deposit. Upon annealing, hardness of the composites increased to ca. 700 HV.

### 3.8. Ni–P electrodeposits reinforced by $CeO_2$

$CeO_2$ is characterized by good corrosion and wear resistance and excellent oxidation performance Its inclusion in the metal matrix is determined to bring an improvement of deposit's mechanical properties, moreover to result in the crystal refinement and a denser structure of the coating.

Jin et al. [105] found Ni-P/$CeO_2$ composites prepared by electroless deposition to have a compact microstructure and good corrosion resistance. Zhou et al. [106] fabricated Ni–P composite coatings containing nanosized $CeO_2$ particles by employing PC electrodeposition under ultrasonic field. Composite coating synthesized by PC plating with 15 g L$^{-1}$ $CeO_2$ in the electrolytic bath exhibited fine-grain structure and improvement in microhardness (575 HV) compared to pure Ni–P coatings fabricated in the absence of $CeO_2$ particles under the same deposition conditions (425 HV). Hardness of the composites was further improved up to 780 HV by annealing at 600 °C for 2 h.

### 4. Ni–P electrodeposits reinforced by solid lubricant particles

It was estimated that 23% of the world's total primary energy consumption originates from tribological contacts [107]. Out of that 20% is on account of overcoming friction and 3% is used to re-manufacture worn components and equipment. Responsible energy management is particularly important, since concerns over global warming grow more important every day and the production of greenhouse gases and the carbon footprint escalate. Hence, improved energy consumption efficiency and sustainable tribology could create a huge impact with the design of new smart solutions which will increase the performance and reduce the systematic energy losses.

Solid lubricant coatings are in high demand owing to the rapid industrial development, where tribological applications involve progressively more extreme settings that involve a vacuum, high temperatures, radiation, oxidative environments, etc. [108,109]. Under such variable and complex operating conditions, liquid and grease-based lubricants cannot provide adequate operation and protection [110,111]. However, solid lubricant coatings allow to overcome many of the issues related to the use of liquids. They exhibit self-lubricating properties, can sustain harsh environments and have the ability to improve machine efficiency and decrease the energy consumption [112]. It has been demonstrated by many authors that the addition of phases possessing lubricant properties (solid lubricants), such as: $MoS_2$, $WS_2$, graphite, hexagonal BN, etc. to a metallic matrix such as Ni–P can result in a significant improvement of tribological properties and the fabrication of self-lubricating coatings.

There are several marked classes of solid lubricant materials: carbon-based materials (graphite, CNT, DLC, nanocrystalline diamond), lamellar solids (TMDs, hexagonal BN), polymers (PTFE), soft metals (Au, Ag, Sn, In), halides and sulfates of alkaline earth metals and finally oxides ($TiO_2$, $B_2O_3$) [109,113].

### 4.1. Ni–P electrodeposits reinforced by PTFE

PTFE has numerous appealing properties, out of which some are: non-stickiness, dry lubricity, low coefficient of friction due to its low intermolecular cohesion, good corrosion resistance, low surface energy, chemical inertness [109,114,115]. However, PTFE is characterized by low thermal conductivity (which can cause failure owing to melting) and poor wear properties [109]. PTFE particles are intrinsically hydrophobic and they aggregate easily in the electroplating solution owing to what it is very difficult to obtain their stable dispersion [115,116].

Not many studies are performed on the inclusion of PTFE particles in the Ni–P metal matrix through electroplating, this composition is dominantly fabricated by the means of auto-catalytic deposition. Losiewicz et al. [117] studied the structure and surface morphology of Ni-P/$TiO_2$-PTFE electrodeposits. They fabricated amorphous Ni–P composites with embedded second phase particles that exhibited a tendency of co-agglomeration. Presence of PTFE was established to cause the reduction of the embedded $TiO_2$ particle mean area and to change coatings surface morphology. In the case of Ni-P/$TiO_2$-PTFE composites greater number of microcracks was observed.

### 4.2. Ni–P electrodeposits reinforced by TMDs

Transition metal dichalcogenides (TMDs) [118–122] have attracted a lot of attention as a second phase in composite coatings designated to efficiently mitigate friction. Good tribological properties of TMDs are associated with their intrinsic structural anisotropy characterized by weak interlayer bonds that allow easy shearing and creation of tribolayer on the counterpart surface [123]. Upon shearing basal plains are determined to reorient parallel to the sliding direction facilitating sliding friction [109,113] in Fig. 13d). This class of solid lubricants is however restricted to use in vacuum or dry nitrogen environment (same friction coefficient however better wear life in dry N$_2$). This is due to their ease of oxidation in the humid atmosphere, the existence of surface irregularities and unsaturated, dangling bonds [124,125]. For instance, when the test environment is switched from dry nitrogen to humid air, the friction coefficient of a typical WS$_2$ film can rise from 0.03 to 0.04 to 0.15–0.20, resulting in a wear life several orders of magnitude shorter [125]. However, studies of Colas et al. [126] have demonstrated that the wear life of MoS$_2$ can be extended if a reasonable amount of contaminants is present in both the coatings and the environment. Chemical reaction with water, carbon and oxygen contamination can help friction and offset the loss of ductile properties of the 3rd body film, hence better wear properties of MoS$_2$ under mild vacuum compared to high vacuum. Additionally, N$_2$ environment is not neutral. Velocity accommodation mechanism is modified by N$_2$ adsorption, resulting in better wear life of MoS$_2$ under dry N$_2$.

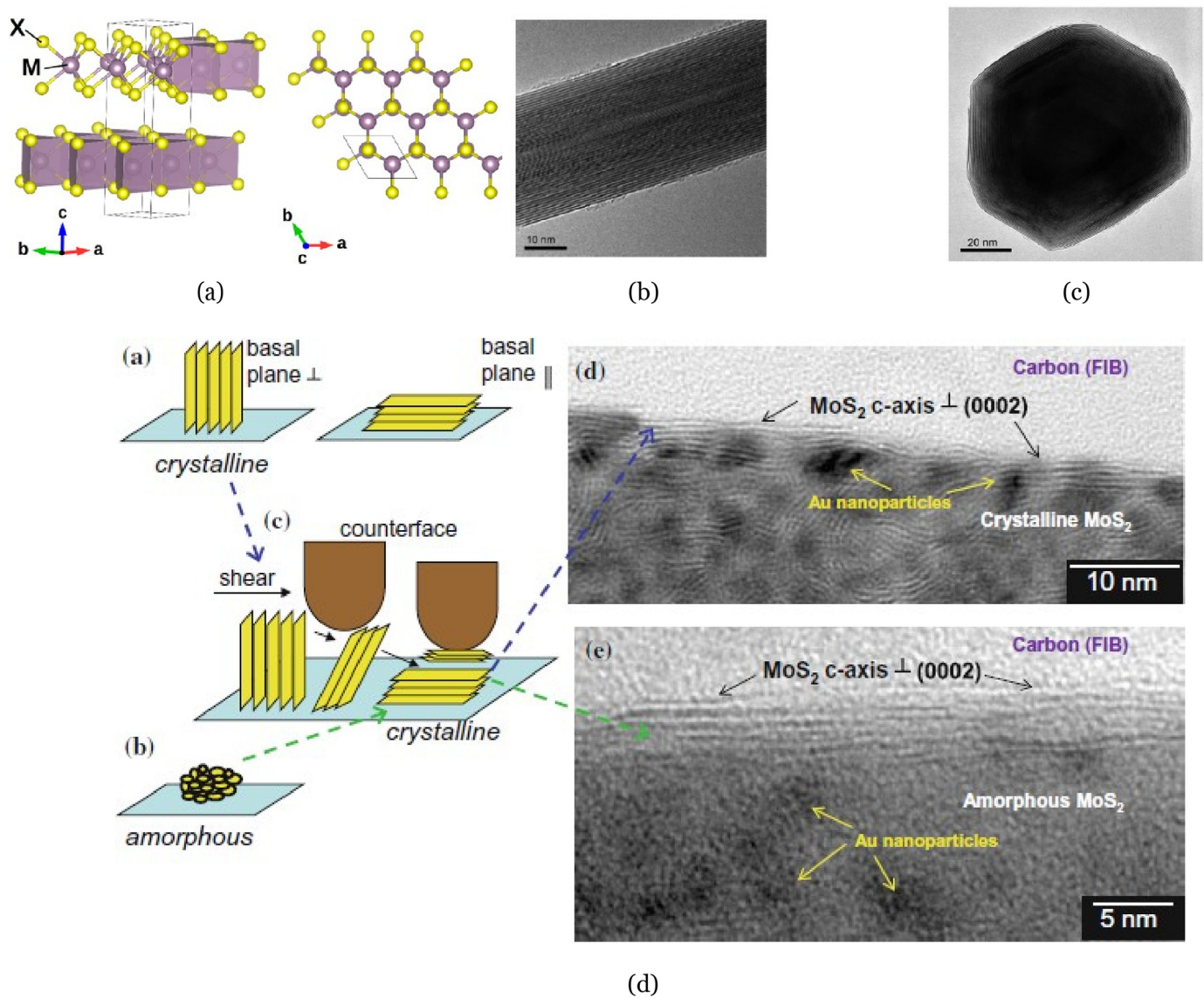

**Fig. 13.** TMD structures and tribological behavior; (a) Hexagonal P6$_3$/mmc structure of 2H polymorph MX$_2$. Reprinted with permission from [152]. Copyright (2015) American Chemical Society; (b) TEM image of a WS$_2$ nanotube. Reprinted with permission from [129]. Copyright (2013) John Wiley and Sons; (c) TEM image of a typical IF-WS$_2$ nanoparticle. Reprinted with permission from [129]. Copyright (2013) John Wiley and Sons.; (d) The process of shear induces reorientation of basal planes parallel to the sliding direction in the case of both crystalline and amorphous structure. Also presented are cross-sectional TEM images inside the wear track of crystalline MoS$_2$/Au and amorphous MoS$_2$/Sb$_2$O$_3$/Au testifying the mechanism. Reprinted with permission from [109]. Copyright (2013) Springer Nature.

It was reported by Tenne and coworkers [127,128] that nanoparticles of organic compounds like TMDs possessing layered structure become unstable in the planar form and are prone to form closed-cage and hollow core nanostructures termed inorganic fullerene-like nanoparticles (IF) and multiwall nanotubes (inorganic nanotubes-INT) in Figs. 13b) and c). These kinds of structures [129,130] are proved to be less prone to oxidation owing to the lack of reactive sights and to be in general more chemically inert [131] compared to conventional 2H-MoS$_2$ and 2H-WS$_2$ hexagonal structures.

Many researchers have postulated that incorporating conventional forms of TMDs or embedding them in a metallic matrix could contribute to their protection against oxidation and could allow to exploit their beneficial tribological properties even in humid atmosphere conditions [132]. In the case of IF structures, the main mechanisms of favourable friction behaviour are postulated to be: rolling friction of the IF nanoparticles which are gradually furnished to the surface from the network of pores of the matrix, IF particles are confined in nanoscopic cavities providing the spacing effect and preventing contact between the asperities and third body material transfer [133]. Due to above mentioned, they are deemed to have even greater potential for solid lubrication applications when compared to conventional structures [134,135].

Incorporation of TMDs in the Ni–P matrix has been mostly achieved through electroless deposition [136,137]. $MoS_2$ and $WS_2$ are two of the most impressive solid lubricants but they a) are very hydrophobic, b) are difficult to disperse in aqueous solutions and c) can give rise to service corrosion in service in some environments, due to local $SO_2$ and $H_2SO_4$ formation. The benefits of such powerful lamellar lubricant particles led He and coworkers to examine codeposition of $WS_2$ [14] and $MoS_2$ [132] in a Ni–P matrix by electroplating. The resultant deposits demonstrated improved tribological and self-cleaning properties.

In their first study, He et al. [132] fabricated Ni-P/$MoS_2$ composite coatings by electrodeposition. Micron size $MoS_2$ particles were co-deposited within the metallic matrix. Friction coefficient (COF) of the coating containing 7.9 wt% $MoS_2$ during sliding against steel in air exhibited a steady-state value of 0.05 during 1 h reciprocating test. The COF value was almost one order of magnitude lower than one of the Ni-P coating without $MoS_2$ inclusion (0.45). Ni-P/$MoS_2$ coatings were characterized by rough surface with nodules. This was, according to the authors, owing to the preferential deposition of Ni atoms on the conductive surfaces of the $MoS_2$ particles once these were attached to the cathode surface. However, with the increase of the $MoS_2$ incorporation composite coatings exhibited much smoother surfaces characterized by the growth of smaller crystals which indicated that the increasing number of incorporated $MoS_2$ particles have provided the increased number of sites for the nucleation of nickel growth. Linear relationship was observed between the particle amount incorporated in the coating and the particle concentration in the bath up to a certain level, above which saturation occurred. $MoS_2$ particles addition caused an increase in hardness despite such particles being softer than the matrix. This was associated with the fact that adsorbed reinforcement particles act as nucleation sights on the cathode, hence they increase charge transfer resistance of the electroplating process resulting in more negative cathode potential and enhance the crystal refinement. The increased number of grain boundaries inhibits the dislocation movement, as per Hall-Petch mechanism.

In a subsequent study, He et al. [14] demonstrated successful incorporation of sub-micron size $WS_2$ particles in the Ni–P matrix via electroplating. According to them, even though surface of pure Ni–P alloy coatings is intrinsically hydrophilic, with a water contact angle of 87°, hydrophobic properties can be obtained by the addition of $WS_2$ particles which additionally impart lubricating properties to the coatings. Sufficient $WS_2$ content in the composite coatings ensured the formation of a uniform tribofilm and provided a very low coefficient of friction (as low as 0.17). Water contact angle for the composite containing 4.8 wt% $WS_2$ was 157°. As in the previous case, the incorporation of $WS_2$ particles into Ni–P metal matrix was found to cause significant changes of surface morphology from: a planar smooth surface, to a nodular, rough surface and eventually to a hierarchical rough surface.

Modification of microstructure, surface morphology and improvement of tribological properties, induced by the inclusion of nano or micron sized $MoS_2$ particles have been observed also in the case of other binary alloys with nickel, such as Ni–Co [138] and Ni–W [139].

### 4.3. Ni–P electrodeposits reinforced by carbon based lubricant particles

Graphite [118] is similar to TMDs in having a layered structure. For graphite to function properly, the presence of water vapour or other gases is necessary (≥100 ppm) resulting in poor tribological properties of graphite in vacuum or dry atmosphere. It has been postulated that these compounds introduce an intercalation or chemisorption effect which weakens the binding force between basal planes near the surface owing to the increase in interlayer spacing, thereby allowing graphite's planes to shear easily. Yen et al. [140] investigated the origin of low-friction behaviour in graphite by surface X-ray diffraction. They demonstrated, within the limits of experimental error, that there was no change in the interlayer spacing of graphite's basal planes near the surface of the outgassed graphite sample exposed to ambient and humid air environments. Such finding supported the alternative explanation for the good tribological properties of graphite, namely that molecules are required to saturate dangling covalent bonds at edge sites of the basal planes for graphite to maintain its low friction behaviour.

Suzuki et al. [141] studied the microstructure and properties of Ni–P matrix composite films with the inclusion of nanosized carbon black particles, produced by electrodeposition. The carbon black content in the deposits increased with increasing its concentration in the bath up to an optimal value reaching a maximum of 0.77 wt%, while phosphorus content was approximately constant at 11.2–13.2 at. %. The composites exhibited higher hardness and better tribological behavior both before and after heat treatment when compared to pure Ni–P alloy.

Since their discovery by Iijima in 1991 [142], carbon nanotubes have attracted much attention, owing to their beneficial properties which include good mechanical characteristics (high tensile strength (60 GPa) and high elastic modulus (1 TPa)), thermal and electrical conductivity, together with favourable chemical and optical properties. Most studies of the tribological properties of the metal-CNT composite coatings have been focused on the electroless nickel plating technique, however several authors have reported the fabrication of Ni-CNT composites through electrodeposition [143–145]. Arai and coworkers [144] fabricated Ni/MWCNT composite electrodeposits having excellent thermal conductivity and improved tribological properties [143]. The friction coefficient of Ni-MWCNT composite films additionally decreased with increasing MWCNT content, minimum value obtained being 0.13 under dry lubrication conditions.

In another study, Suzuki and colleagues [146] investigated microstructure, mechanical and tribological properties of Ni-P/MWCNT electrodeposits. They fabricated composite films containing 20–22 at.% of P and 0.7–1.2 wt% of MWCNT. Obtained deposits exhibited higher hardness both before and after heat treatment and lower friction coefficient when compared with pure Ni–P alloy films. The friction coefficient of the Ni–P alloy film against steel gradually increased at higher cycle numbers. In contrast, the friction coefficient of the Ni-P/MWCNT composite films decreased rapidly at an early stage and reached a steady value, indicating solid lubricity caused by the intrinsic solid lubricity of the MWCNTs. At 50 cycles, the friction coefficient of the Ni-P/MWCNT composite film was approximately 0.1–0.2, both before and after heat-treatment.

Wang et al. [147] investigated structural, mechanical and tribological properties of Ni-P/MWCNT coatings annealed at temperatures from 350 °C to 500 °C. MWCNT incorporation amount was low (1.9 wt%). Results indicated that the annealed coatings comprised a hard $Ni_3P$ phase and consequently presented a higher hardness of approx. 7–8 GPa than the as-plated samples at approx. 6 GPa. Maximum hardness was obtained after annealing at 350 °C. Annealed samples exhibited lower friction coefficients (0.71–0.86) compared to as-plated coatings (0.87). This effect was attributed to the formation of $H_3PO_4$ in the course of the tribochemical reaction of $Ni_3P$ with ambient environment. Lower wear rate was obtained after annealing at temperatures < 380 °C, owing to the decomposition of the amorphous carbon in MWCNT above this temperature.

### 4.4. Ni–P electrodeposits reinforced by hexagonal BN

Boron nitride with a hexagonal close-packed structure (hBN) possesses the same crystalline organization as graphite having the same total number of electrons available for bonding [118]. However, adsorption of atmospheric gases in the case of hBN does not produce an amelioration of its lubricating properties. This is attributed to the lack of unpaired electrons present in graphite and their pairing in p orbitals in hBN.

Data published on the incorporation of hBN in the Ni–P matrix refer to the deposition of composite coatings mostly by electroless plating

[148,149]. The findings showed an improvement in wear resistance and coefficient of friction for the fabricated composites. Peng and coworkers [150] studied the effect of incorporation of 2 μm hBN particles in the Ni−P matrix through PC electrodeposition. Composite coatings were characterized by an amorphous structure and demonstrated a significant improvement in friction coefficient and wear resistance with a slight decrease in hardness compared to Ni−P coatings. The lowest coefficient of friction experienced was 0.08 in the case of the bath containing the highest content of hexagonal BN (20 g L$^{-1}$). There are no studies of the influence of hexagonal BN inclusions on the corrosion properties of Ni−P alloys. Studies on pure Ni deposits [151] produced from the bath containing 5–20 g L$^{-1}$ h-BN nanosheets indicated increased microhardness, refined surface morphology and improved corrosion protection in 3.5 wt% NaCl solution.

## 5. Conclusions

The study of Ni−P composites produced by electrodeposition still lags behind the advances made on the investigation of composites produced by electroless deposition, even though research demonstrates that the codeposition of particles in the Ni−P matrix by the means of electroplating can be optimised to meet the increasing requirements on process simplicity and obtained deposits multi-functionality. Insufficient attention is paid to the investigation of bath ageing, incorporation combining or sequencing different deposition techniques. It is important to iterate that obtaining a stable dispersion of non-agglomerated reinforcement particles is of critical significance and further investigations of the possibilities and approaches to interrogate and control this aspect are necessary. Influence of hydrodynamics and mass transport need to be further investigated especially at the larger scale when the influence of the variability of all relevant conditions and parameters is even more pronounced and produces a complex combination effect on deposit structure and properties. Current distribution is also vital but often its influence is poorly defined. Despite small advances, modelling of composite electrodeposition remains unsatisfactory. Finally, there is a large gap between scientific research and industrial needs. So, research should continue to bridge this gap in order to progress industrial processing, scale-up and technological development.

## 6. Areas needing future research and development

In order to realize rapid developments in the science and technology of Ni−P composite coatings containing well dispersed included particles, it is important to carry out further R & D on several critical (and interactive) aspects which are poorly appreciated in the literature:

1. *Adequate preparation of particle suspensions in the bath*; this may necessitate high shear mixing of a particle in bath slurry, followed by ultrasonication of the electrolyte. The controlled use of surfactants must be considered, including their type (cationic, anionic, nonionic or mixed), concentration, environmental impact and effect on deposit properties.
2. *Maintenance of an effective particle dispersion in the bath*, using adequate electrode/electrolyte movement and electrolyte agitation to provide well-defined electrode geometry, a known flow regime (laminar or turbulent) and controlled fluid flow.
3. *Characterisation of the bath quality*, including its conductivity, throwing power, stability of particle suspension, seta potential of particles, particle size distribution and particle sedimentation time.
4. *An improved appreciation* of the rate control of electrode processes and transport of species, together with the particle incorporation and consolidation processes in deposition.
5. *Improved computational models* which quantitatively relate bath composition and operating conditions to deposit properties.
6. *Considered experimental designs* which facilitate a systematic examination of parameters via an established algorithm, allowing optimisation of the results.
7. *Long-term ageing effects in the bath* on the deposit quality, since electrolyte additives can decompose at anodes and cathodes or during prologed heating.
8. *Sufficient attention to the parameters affecting facile scale-up* and industrial scale processing, including electrode geometry and size, interelectrode gap, electrolytic power required, bath volume, bath composition, capital and running costs. Particular attention must be paid to the feasibility of large scale ultrasonication of baths and the availability, reliability and cost of high current pulsed current power supplies.
9. *Performance, cost and convenience based comparisons* between Ni−P based coatings achieved by various deposition techniques (e.g., electrodeposition, electroless deposition and vacuum deposition) and between Ni−P-particle composites and competitive tribological coatings (for a particular application).
10. *The continuing fashion for use of nanoparticles needs to give way to a more balanced appraisal of their use*, in terms of availability, cost, ease of handling, environmental consequences and relative performance vs. conventional micron sized inclusions.
11. *The prevalence of an experimental electrodeposition geometry consisting of parallel vertical electrodes in a magnetically stirred beaker* is understandable due to its convenience. However, it should be realised that the size of the interelectrode gap is critical, as are the material, diameter, length and shape of the stirrer follower, not just its rotation speed. Moreover, this type of stirring leads to poorly defined flow and mass transport rates. Hence, little scale-up information is achieved from the use of this arrangement.
12. *The combination of hard particles to improve dispersion hardening and wear resistance of a composite deposit and lamellar solid state lubricants* could be used to provide a new generation of tribological coatings.

## Acknowledgements

This work was supported by the European Union's Horizon 2020 research and innovation programme SOLUTION, under grant agreement No. 721642. FCW is grateful to Ir Kees Helle, who acted as an excellent industrial supervisor and mentor during an early, undergraduate research project on electrodeposited nickel composite coatings containing diamond or PTFE inclusions at AKZO R & D laboratories, Arnhem, The Netherlands, in 1974.